\pdfoutput=1
\documentclass{ieeeaccess}
\usepackage{cite}
\usepackage{amsmath,amssymb,amsfonts}
\usepackage{graphicx}
\usepackage{textcomp}
\usepackage{booktabs}
\usepackage{multirow}
\usepackage{xcolor}
\usepackage{hyperref}
\providecommand{\xfigwd}{0pt}
\newcommand{\orcidicon}[1]{\,\href{https://orcid.org/#1}{\textcolor[HTML]{A6CE39}{\scriptsize ORCID}}}

\begin{document}
\vol{14}\year{2026}\setcounter{page}{159620}

\history{Date of publication October 15, 2026.}
\doi{10.1109/ACCESS.2026.3743502}

\title{The Fusion Tax: A Diagnostic Analysis of When Hybrid Retrieval Hurts}

\author{\uppercase{Aaditya Chauhan}\,\orcidicon{0009-0000-8652-0071}\authorrefmark{1}
and \uppercase{Arun Agarwal}\,\orcidicon{0009-0005-3065-9804}\authorrefmark{2}}

\address[1]{Independent Researcher, Bothell, WA 98012 USA (e-mail: aadityc28@gmail.com)}
\address[2]{Independent Researcher, Issaquah, WA 98027 USA (e-mail: arunagar28@gmail.com)}

\tfootnote{This work received no external funding. The authors declare no conflict of interest. This is the author-prepared version of the article published in \textit{IEEE Access}, vol.~14, pp.~159620--159637, 2026, DOI: \href{https://doi.org/10.1109/ACCESS.2026.3743502}{10.1109/ACCESS.2026.3743502}, under a Creative Commons Attribution 4.0 (CC BY 4.0) license.}

\markboth{Chauhan and Agarwal: The Fusion Tax: A Diagnostic Analysis of When Hybrid Retrieval Hurts}
{Chauhan and Agarwal: The Fusion Tax: A Diagnostic Analysis of When Hybrid Retrieval Hurts}

\corresp{Corresponding author: Aaditya Chauhan (e-mail: aadityc28@gmail.com).}

\begin{abstract}
Hybrid retrieval, which fuses a sparse lexical retriever with a dense neural one, is widely recommended in the software documentation surveyed in September 2026, usually with reciprocal rank fusion (RRF), on the strength of aggregate gains. We ask what fusion costs on individual queries, treating ``do not fuse'' as a first-class outcome, in the setting in which hybrid retrieval is deployed: a standard Anserini BM25, five dense encoders spanning 22M--596M parameters and three model generations, a learned-sparse (SPLADE) alternative, and a cross-encoder second stage, on seven BEIR datasets and three fusion functions. Even the best method, dense-weighted score interpolation, ranks relevant documents below the better of its two components on 13--48\% of queries and below the deployable dense-only retriever on 8--21\% (14--48\% and 11--25\% at a weight selected under leave-one-corpus-out validation), while beating dense-only in aggregate on every corpus with the primary encoder; the downside is significant under loss-averse risk-sensitive evaluation on six of the seven corpora. We call this loss the fusion tax. It is mostly a ranking effect that preserves recall in aggregate, it is most likely on the queries that only one retriever covers, and it persists across the encoder ladder and the SPLADE substitution. RRF carries the higher tax wherever the dense retriever is clearly stronger and only beats dense-only where the two components are balanced. Fusion failure is weakly predictable within a corpus but transfers unreliably across corpora, because the features that predict it change sign with the sparse--dense balance. A cross-encoder recovers part of the tax only where it is itself reliable. We distill these findings into guidance that keys the fusion decision to the measured sparse--dense balance of the target corpus.
\end{abstract}


\titlepgskip=-15pt

\maketitle

\section{Introduction}
\label{sec:intro}

Hybrid retrieval, which fuses a sparse lexical retriever such as BM25~\cite{robertson2009bm25} with a dense neural retriever~\cite{karpukhin2020dpr}, is, at the time of writing, the default recommendation of most current search stacks and retrieval-augmented generation (RAG) frameworks, and reciprocal rank fusion (RRF)~\cite{cormack2009rrf} is the fusion method most of them recommend or ship as the only option (Section~\ref{sec:related-fusion}). The recommendation rests on aggregate evidence: averaged over a query set, the fused ranking usually beats either component.

That fusion can help on average while hurting individual queries is not a new observation. Risk-sensitive evaluation of rank fusion~\cite{benham2017riskreward}, fusion-oriented query performance prediction~\cite{markovits2012predicting,roitman2018enhanced}, and per-query routing between retrievers~\cite{arabzadeh2021predicting,mothe2023selective,hsu2025dat} all start from it. What those lines do not provide is the quantity a practitioner deciding between a hybrid and a single strong retriever needs: how often, and by how much, a modern sparse--dense fusion ranks relevant documents \emph{below the better of its own two components}, and below the single component one would deploy instead. We call this per-query loss the \textbf{fusion tax}, we treat ``do not fuse'' as a first-class outcome, and we measure the tax in the setting in which hybrid retrieval is actually deployed today: a standard BM25 pipeline, dense encoders spanning 22M--596M parameters and three model generations, a learned-sparse alternative, and a cross-encoder second stage.

Concretely, this paper contributes:
\begin{enumerate}
    \item A \textbf{per-query accounting of fusion} on seven BEIR corpora and three fusion functions that reports harm and benefit symmetrically, against a per-query oracle and against the deployable always-dense baseline, overall and by a four-quadrant diagnostic taxonomy (Sections~\ref{sec:taxonomy-results}--\ref{sec:fusion-tax-results}). With a standard Anserini BM25 and the best fusion method, dense-weighted interpolation, 13--48\% of queries pay an oracle tax and 8--21\% score below dense-only, even though fusion beats dense-only in aggregate on every corpus with the primary encoder; the loss is significant under loss-averse risk-sensitive tests on six of the seven corpora.
    \item A \textbf{held-out selection of the interpolation weight} under leave-one-corpus-out and within-corpus cross-validation (Section~\ref{sec:alpha}), so that the recommendation of a dense-heavy weight, and the tax at that weight, are validated on queries not used to choose it.
    \item A \textbf{robustness analysis on all seven corpora} that falsifies the ``stronger retrieval dissolves the tax'' hypothesis: a controlled encoder scale ladder (BGE-small/base/large, plus MiniLM and the 2025-generation Qwen3-Embedding) leaves the tax at 14--43\% for the newest and largest encoder, a range overlapping those of every smaller and older one (Section~\ref{sec:ladder}), and replacing BM25 with SPLADE shows that interpolation's tax is intrinsic to score fusion, whereas RRF's tax responds to the quality of the lexical component (Section~\ref{sec:splade}).
    \item A \textbf{predictability analysis with a negative cross-corpus result}: fusion failure is weakly predictable from cheap query features within a corpus (mean held-out AUC 0.63) but transfers unreliably across corpora (mean 0.59, ranging from chance at 0.49 to 0.74), a held-out fusion/dense-only router built on the same features never beats static interpolation, and a feature-level analysis shows why, since both the important features and the direction of their association with failure change from corpus to corpus (Section~\ref{sec:feature-results}).
    \item A \textbf{measurement of downstream exposure}: on the three corpora where a strong cross-encoder over budget-matched candidate sets improves every retrieval arm, the prevalence of the deployable tax falls by a third on one and not at all on the other two, while 71--90\% of the originally hurt queries are individually repaired and others take their place; where it degrades or overrides the first stage it overwrites the tax rather than repairing it; once reranked, interpolated and dense candidate sets differ by at most 0.011 nDCG@10 with the one reranker and depth tested (paired intervals within $\pm0.02$) except where sparse and dense are balanced (Section~\ref{sec:rerank-results}), and measured index sizes and latencies quantify what the second retriever costs (Section~\ref{sec:efficiency}).
\end{enumerate}

\noindent\textbf{Scope.} Our object of study is first-stage retrieval quality, measured with standard ranking metrics, plus the recovery obtained from one reranking stage. We do not evaluate answer generation. Because the tax is largely a ranking-position effect (recall@10 is preserved in aggregate, although a minority of the hurt queries do lose a relevant document from the top-10; Section~\ref{sec:fusion-tax-results}), its consequence for a downstream consumer depends on how rank-sensitive that consumer is; the reranking experiment measures the case that matters most in practice, and we state the remaining exposure without claiming an effect on generated answers.

\section{Related Work}
\label{sec:related}

\subsection{Dense and Sparse Retrieval}
Dense retrievers such as DPR~\cite{karpukhin2020dpr} and late-interaction models such as ColBERT~\cite{khattab2020colbert} match queries and documents in an embedding space and therefore handle paraphrase and synonymy, but can miss rare terms, identifiers, and out-of-distribution vocabulary. BM25~\cite{robertson2009bm25} remains a strong and cheap baseline, especially for keyword- and entity-heavy queries, and its Anserini/Pyserini implementation~\cite{yang2018anserini,lin2021pyserini} is the reference point for BEIR~\cite{thakur2021beir,kamalloo2024brewingbeir}. Learned sparse models such as SPLADE~\cite{formal2021splade,formal2022splade} add learned term expansion to an inverted-index representation. Because each family fails on a \emph{different} slice of queries, hybrid fusion is attractive; for the same reason, aggregate gains can coexist with per-query harm.

\subsection{Hybrid Fusion and the RRF Default}
\label{sec:related-fusion}
Hybrid systems combine sparse and dense rankings either by rank or by score. RRF~\cite{cormack2009rrf} sums reciprocal ranks with a constant $k$: it is parameter-light and scale-free, but discards score magnitude, treating a confidently top-ranked document identically to a marginal one. Score-based methods (CombSUM/CombMNZ~\cite{fox1994combmnz,lee1997analyses} and weighted interpolation~\cite{vogt1999linear}) preserve magnitude but must first place two incomparable score distributions on a common scale.

A review of vendor and framework documentation, as accessed in September 2026, shows that RRF is the recommended or sole fusion method in Elasticsearch~\cite{elastic_hybrid_docs,elastic_rrf_docs}, Azure AI Search~\cite{azure_rrf_docs}, and LangChain's \texttt{EnsembleRetriever}~\cite{langchain_ensemble}; Qdrant documents it as the safe default~\cite{qdrant_hybrid_docs}, Pinecone recommends it when fusing across indexes~\cite{pinecone_hybrid_docs}, and it is the fusion used in the canonical hybrid examples of Milvus~\cite{milvus_rerank_docs} and LlamaIndex~\cite{llamaindex_fusion}. Recent RAG systems adopt it for the same reason, explicitly to avoid score calibration between heterogeneous retrievers: the winning SIGIR 2025 LiveRAG system fuses BM25 and dense lists with RRF ($k{=}60$)~\cite{ran2025rmitliverag} in a challenge where most teams used hybrid search~\cite{carmel2025liverag}, as do RAG-Fusion~\cite{rackauckas2024ragfusion} and several 2025--2026 systems~\cite{meng2026sifei,parames2026fragata,mala2025hybridhallucination}. The default is not universal: OpenSearch's hybrid pipeline defaults to min--max normalization with an arithmetic mean and added RRF only in version 2.19~\cite{opensearch_hybrid_docs}, and Weaviate reversed its default from rank fusion to min--max ``relative score'' fusion in version 1.24 after internal benchmarks~\cite{weaviate_hybrid_docs,weaviate_fusion_blog}; Vespa's tutorial reports linear normalization beating RRF on NFCorpus~\cite{vespa_hybrid_docs}. Elasticsearch and OpenSearch, the two largest open-source engines, thus disagree on the default, which is exactly the rank-versus-score question our per-query analysis speaks to. This survey is a time-sensitive account of documented defaults rather than a peer-reviewed comparison of effectiveness: it rests on documentation that changes from release to release (the OpenSearch and Weaviate changes above are two examples), and it records what practitioners are steered toward, not what has been shown to work; the peer-reviewed comparisons of rank and score fusion are reviewed next.

\subsection{Score Calibration versus Rank Fusion}
\label{sec:related-calibration}
When is score fusion preferable to rank fusion? The metasearch literature answered this early: Lee~\cite{lee1997analyses} attributed fusion gains to the ``unequal overlap'' of relevant versus non-relevant documents across runs, Vogt and Cottrell~\cite{vogt1999linear} characterized when a linear combination of scores helps, and Renda and Straccia~\cite{renda2003rankvsscore} found score-based aggregation superior to rank-based aggregation whenever scores are available and can be normalized. The normalization step is the crux: simple min--max or z-score normalization~\cite{montague2001normalization} is sensitive to outliers and to the depth of the list, whereas score-distribution models~\cite{manmatha2001score,arampatzis2009snr} calibrate scores on a relevance scale at the cost of fitting a model per run, and the interpolation weight itself can be set from component effectiveness~\cite{wu2009weights}. For sparse--dense hybrids, Ma et al.~\cite{ma2021replicationdpr} and Chen et al.~\cite{chen2022oodhybrid} used weighted score interpolation and reported consistent zero-shot gains, and Bruch et al.~\cite{bruch2023fusion} showed that a convex combination of min--max normalized scores outperforms RRF and is sample-efficient to tune. The practical reading is that interpolation wins when the two score distributions can be put on a comparable scale and one component is reliably stronger, because it keeps the stronger component's confidence; rank fusion is the safer choice when scores are incomparable or unavailable, or when the two components are similarly strong, so that neither should dominate. Our results reproduce both halves of this picture (Section~\ref{sec:fusion-results}).

\subsection{Query Performance Prediction}
Query Performance Prediction (QPP) estimates retrieval quality without relevance judgments~\cite{shtok2012qpp}. Post-retrieval predictors are computed from the score distribution of a single ranking: NQC~\cite{shtok2012qpp} measures the standard deviation of top scores and WIG~\cite{zhou2007wig} the gain of the top results over the collection mean; both are free, since the retrieval scores are already in hand. Faggioli et al.~\cite{faggioli2023qppneural} report that classical predictors degrade on neural rankers, which is one reason cheap routing over dense retrievers is hard. Recent work applies QPP to \emph{weight} the components in fusion~\cite{santra2026qpp}. We ask instead whether the same signals predict fusion \emph{failure}, and whether that prediction transfers across corpora.

\subsection{Selective Retrieval and Per-Query Routing}
\label{sec:related-routing}
Routing studies differ in what the per-query decision is. Some decide \emph{whether} to retrieve at all~\cite{mallen2023whennot}; some choose \emph{which} retriever or pipeline to run, from sparse versus dense on efficiency grounds~\cite{arabzadeh2021predicting} to a mixture of expert embedding models~\cite{lee2025routerretriever} or a complexity-tiered pipeline~\cite{jeong2024adaptiverag,hashmi2026adaptive}; and a growing group sets the \emph{fusion weight} per query, from an LLM judge of each component's likely effectiveness (DAT~\cite{hsu2025dat}) to query-specificity heuristics~\cite{mala2025hybridhallucination}, zero-shot mixtures of sparse and dense retrievers~\cite{kalra2025mor}, margin-based confidence~\cite{kim2026qudar}, or a learned regressor from the query representation~\cite{posokhov2026queryadaptive}. Mothe and Ullah~\cite{mothe2023selective} frame configuration selection as a risk-sensitive decision. All of these presume that per-query component effectiveness is reliably learnable. Our work is complementary and diagnostic: rather than proposing another router, we characterize when fusion fails and test whether the cheap signals such routers rely on transfer across corpora, finding that they do not (Section~\ref{sec:feature-results}). This also suggests why DAT resorts to an LLM judge: the signal it needs is not available from cheap, transferable query features.

\subsection{Risk-Sensitive Fusion and Fusion-QPP}
\label{sec:related-risk}
That fusion can harm individual queries while helping on average has roots in risk-sensitive evaluation~\cite{wang2012robust,dincer2014trisk}. Benham and Culpepper~\cite{benham2017riskreward} gave a per-query risk--reward view of rank fusion, reporting that unsupervised fusion improves most topics but degrades a minority (e.g.\ 142 improved vs.\ 21 worsened on Robust04). The fusion-QPP line~\cite{markovits2012predicting,roitman2018enhanced} predicts the absolute effectiveness of an already-fused list from reference lists; because it never compares fusion to a component, it cannot establish that fusion \emph{hurts}, and because its estimators are fit per collection, it makes no cross-corpus claim. Table~\ref{tab:contrib} positions our study against these lines. The fusion tax descends from the risk framing but differs in the \emph{reference}: wins and losses in~\cite{benham2017riskreward} are scored against a single weak BM25 run with a $\pm10\%$ deadband, so a query can be a win there (fusion beats weak BM25) yet a loss for us (fusion below dense-only). We score loss against the better component per query and against the deployable always-dense system, in a modern sparse--dense setting, across an encoder scale ladder and a learned-sparse substitution, under held-out weight selection, and we test whether the loss is predictable across corpora and recoverable by reranking. Each prior line covers at most a few of these axes, and none measures the loss against the better component or against the deployable single retriever; two concurrent aggregate-level studies~\cite{louis2024knowfuse,wang2025balancing} corroborate the phenomenon without measuring it per query.

\begin{table*}[!t]
\caption{Positioning against prior lines of work. ``Loss reference'' is the quantity a per-query loss is measured against; LOCO = leave-one-corpus-out.}
\label{tab:contrib}
\centering
{\scriptsize\setlength{\tabcolsep}{3pt}\renewcommand{\arraystretch}{1.1}%
\begin{tabular}{@{}p{2.55cm} p{2.15cm} p{2.35cm} p{2.25cm} p{2.35cm} p{2.25cm} p{2.9cm}@{}}
\toprule
 & Bruch et al.~\cite{bruch2023fusion} & Benham \& Culpepper~\cite{benham2017riskreward} & Fusion-QPP~\cite{markovits2012predicting,roitman2018enhanced} & Routing~\cite{arabzadeh2021predicting,mothe2023selective,hsu2025dat} & Concurrent~\cite{louis2024knowfuse,wang2025balancing} & This work \\
\midrule
Unit of analysis & aggregate & per-query & per-query & per-query & aggregate & per-query \\
Loss reference & -- & fixed weak BM25 run & absolute fused effectiveness & -- & best single system (mean) & $\max$(BM25, dense) and always-dense \\
Components & sparse + dense & classical lexical runs & classical lexical runs & sparse / dense / hybrid & sparse + dense & sparse + dense; SPLADE + dense \\
Fusion functions & RRF, convex, learned & RRF, CombSUM, \ldots & CombSUM/MNZ & convex ($\alpha$) & RRF, convex & RRF, convex, CombMNZ \\
Dense-encoder scale ladder & -- & -- & -- & -- & -- & \checkmark\ 22M--596M, three generations \\
Learned-sparse substitution & \checkmark & -- & -- & -- & -- & \checkmark\ (per-component attribution) \\
Held-out fusion weight & \checkmark & -- & -- & \checkmark & partly & \checkmark\ (LOCO + CV) \\
Predicts per-query failure & -- & -- & \checkmark\ (absolute) & \checkmark & -- & \checkmark \\
Cross-corpus transfer tested & -- & -- & -- & -- & -- & \checkmark\ (negative result) \\
Risk-sensitive significance & -- & \checkmark & -- & \checkmark & -- & \checkmark\ (TRisk) \\
Reranking recovery measured & -- & -- & -- & -- & -- & \checkmark \\
\bottomrule
\end{tabular}}

\end{table*}

\subsection{First-Stage Fusion, Reranking, and RAG}
\label{sec:related-rerank}
Production systems rarely serve the first-stage list directly: a cross-encoder or LLM reranker re-sorts the top candidates~\cite{lin2021pretrained,schlatt2025rankdistillm,abdallah2025howgood,pandit2025rerankingsurvey}, and RAG pipelines then pass a truncated context to a reader. Reranking does not make the first stage irrelevant. Rerankers are conditioned on the candidate distribution they were trained on~\cite{gao2021rethink,lu2022hyrr}, their effectiveness degrades as the candidate pool grows~\cite{jacob2025drowning}, and the right depth is itself query dependent~\cite{meng2024rlt}; an industrial study of RRF-based RAG-Fusion found its recall gains ``largely neutralized'' after reranking and truncation~\cite{medrano2026ragfusion}, and the TREC 2024 RAG baselines use BM25 plus a reranker with no fusion at all~\cite{pradeep2025ragnarok}. On the reader side, answer quality depends on where relevant passages sit in the context~\cite{liu2024lostmiddle,hsieh2024foundmiddle} and on the composition of the retrieved set~\cite{cuconasu2024noise,mazuryk2026powerless,hsia2025ragged}, which is why per-query retrieval utility for generation is evaluated in its own right~\cite{salemi2024erag}. We do not evaluate generation. We measure the tax at the first stage, and we measure directly how much of it a strong cross-encoder recovers (Section~\ref{sec:rerank-results}), which measures the downstream exposure of rank-sensitive consumers at one reranking depth without claiming an effect on answers.

\section{Problem Formulation and Terminology}
\label{sec:problem}

We fix the vocabulary used in the rest of the paper here; each term is defined once and used verbatim afterwards.

\subsection{Diagnostic Taxonomy}
\label{sec:taxonomy}
For a query $q$, let $\text{NDCG}_s(q)$ and $\text{NDCG}_d(q)$ denote the nDCG@10 of the sparse and dense retrievers. We classify $q$ into one of four \textbf{quadrants} by whether each retriever places at least one relevant document in its top-10:
\begin{itemize}
    \item \textbf{Both-Win}: $\text{NDCG}_s(q) > 0$ and $\text{NDCG}_d(q) > 0$;
    \item \textbf{Dense-Only}: $\text{NDCG}_d(q) > 0$ and $\text{NDCG}_s(q) = 0$;
    \item \textbf{Sparse-Only}: $\text{NDCG}_s(q) > 0$ and $\text{NDCG}_d(q) = 0$;
    \item \textbf{Both-Fail}: $\text{NDCG}_s(q) = 0$ and $\text{NDCG}_d(q) = 0$.
\end{itemize}
The criterion is deliberately coarse: it records \emph{which} retriever covers a query, not how well. Magnitudes are carried by the tax below.

\subsection{The Fusion Tax}
\label{sec:fusion-tax}
For a fused ranking with $\text{NDCG}_f(q)$, the \textbf{fusion tax} of query $q$ is
\begin{equation}
\tau(q) \;=\; \max\!\big(\text{NDCG}_s(q),\; \text{NDCG}_d(q)\big) \;-\; \text{NDCG}_f(q).
\label{eq:tax}
\end{equation}
Fusion \textbf{hurts} $q$ when $\tau(q) > 0$ and \textbf{helps} it when $\tau(q) < 0$; this sign convention (positive is harm) is used in every table and figure. Because BEIR corpora carry multiple, often graded, relevant documents per query, a positive tax means that fusion demoted relevant documents relative to the better component, which may be the loss of the single relevant document or a reordering among several; we therefore speak of relevant documents rather than of a correct answer.

The reference in Eq.~\ref{eq:tax} is a per-query \textbf{oracle}: it presumes knowledge of which retriever wins each query, and Section~\ref{sec:feature-results} shows that neither fusion failure nor the dense-versus-fusion decision is predictable well enough from cheap features for a router to exploit. We therefore report two quantities. The \textbf{oracle tax} is the fraction of queries with $\tau(q) > 0$, an upper bound on per-query harm. The \textbf{deployable tax} replaces the oracle by the best \emph{static} single retriever, always-dense (the stronger component on every corpus with the primary encoder, BGE-small): it is the fraction of queries on which fusion scores below dense-only, i.e.\ the harm a practitioner who simply chose dense retrieval would avoid. Both are accompanied by the symmetric \textbf{benefit} statistics (fraction of queries helped and mean gain when helped) and by the \textbf{net} change, so that harm is never reported without the corresponding benefit. For a fusion method $f$ and baseline $b$ the net mean change decomposes as
\begin{equation}
\mathbb{E}[\Delta] \;=\; P(\text{help})\,\mathbb{E}[\text{gain}\mid\text{help}] \;-\; P(\text{hurt})\,\mathbb{E}[\text{loss}\mid\text{hurt}],
\label{eq:net}
\end{equation}
which is also the reward and risk decomposition of URisk~\cite{wang2012robust}; the risk-sensitive tests in Section~\ref{sec:fusion-tax-results} weight the second term by $(1{+}\alpha_{\text{risk}})$.

\section{Experimental Setup}
\label{sec:setup}

\subsection{Datasets}
We evaluate on seven BEIR benchmark datasets spanning scientific-claim, financial, medical, argumentative, scientific-citation, biomedical, and duplicate-question domains (Table~\ref{tab:datasets}). Every analysis in this paper, including the encoder scale ladder and the SPLADE substitution, uses all seven. For ArguAna, where every query is also a corpus document, we follow the BEIR protocol and remove the query's own document from every retriever's ranking before fusion and evaluation.
\begin{table}[t]
\caption{Dataset statistics (BEIR test splits).}
\label{tab:datasets}
\centering
\begin{tabular}{llrr}
\toprule
Dataset & Domain & \# Docs & \# Queries \\
\midrule
SciFact & Scientific claims & 5,183 & 300 \\
FiQA & Financial QA & 57,638 & 648 \\
NFCorpus & Medical/nutrition & 3,633 & 323 \\
TREC-COVID & Biomedical & 171,332 & 50 \\
ArguAna & Argument retrieval & 8,674 & 1,406 \\
SciDocs & Scientific citations & 25,657 & 1,000 \\
Quora & Duplicate questions & 522,931 & 10,000 \\
\bottomrule
\end{tabular}

\end{table}

\subsection{Retrieval Methods}
\label{sec:retrievers}
\textbf{Sparse.} BM25 is the Anserini implementation~\cite{yang2018anserini} run through Pyserini~\cite{lin2021pyserini} (version 1.6, Java 21) over the official prebuilt BEIR Lucene indexes~\cite{kamalloo2024brewingbeir}, in the multifield configuration (title and body indexed as separate fields) with Anserini's default parameters $k_1=0.9$, $b=0.4$. Section~\ref{sec:baseline-validation} shows that these runs reproduce the published Anserini nDCG@10 on all seven corpora to four decimals. The single-field (``flat'') Anserini index and the Python \texttt{rank\_bm25} implementation~\cite{brown2022rankbm25} serve only as an implementation-sensitivity check (Supplementary Table~S1).

\textbf{Dense.} We use five embedding models spanning 22M--596M parameters and three model generations (2021, 2023, 2025): all-MiniLM-L6-v2~\cite{reimers2019sentbert} (22M, 384-d), BGE-small/base/large-en-v1.5~\cite{xiao2023bge} (33M/110M/335M), and Qwen3-Embedding-0.6B~\cite{zhang2025qwen3embedding} (596M, 1024-d, released 2025). All corpora are searched with exact inner product (FAISS~\cite{johnson2021faiss}). Detailed per-query analysis uses BGE-small; the BGE ladder is the controlled same-family scale-up, MiniLM the cross-family check, and Qwen3 the newer-generation check. Queries are encoded prefix-free with the BGE models throughout; Supplementary Table~S8 re-encodes the queries with the model-card instruction and shows that the tax is insensitive to the choice: the instruction raises dense nDCG@10 by at most 0.026 on five corpora (under 0.02 in all but one model--corpus pair), by 0.09--0.11 on TREC-COVID, and by 0.06 for BGE-base and BGE-large on Quora, and it moves the interpolation tax by at most three points on the five corpora other than TREC-COVID and Quora, where it removes the five-point penalty that prefix-free encoding imposes on BGE-base and BGE-large. Qwen3-Embedding is instruction-tuned and is used with the query instruction prescribed by its model card, with documents truncated at 512 tokens like the other encoders.

\textbf{Learned sparse.} SPLADE~\cite{formal2021splade} (the CoCondenser-EnsembleDistil checkpoint~\cite{formal2022splade}) replaces BM25 as the sparse side in Section~\ref{sec:splade}, scored by exact sparse dot product.

\textbf{Fusion.} RRF with the literature default $k=60$; normalized score interpolation, $s_f = \alpha\,\tilde{s}_d + (1-\alpha)\,\tilde{s}_s$ with min--max normalization over each retriever's top-100 and dense weight $\alpha$; and CombMNZ~\cite{fox1994combmnz}. The interpolation weight is held fixed at $\alpha=0.7$ for every encoder and corpus in every table unless stated otherwise, so no encoder or corpus is tuned within this study; Section~\ref{sec:alpha} selects $\alpha$ under two held-out protocols and reports the tax at the selected weights. All retrievers return the top-100 documents per query, and scores and ranks are stored per query. Exact fused-score ties, frequent under RRF because its contributions take finitely many values, are broken by document identifier in the trec\_eval convention, so fused rankings and the candidate lists of Section~\ref{sec:setup-rerank} are deterministic.

\subsection{Reranking and Efficiency Measurements}
\label{sec:setup-rerank}
To measure how much of the tax a second stage recovers, we rerank with the cross-encoder BGE-reranker-v2-m3~\cite{chen2024bgem3}. For each query, the top-30 candidates of BM25, dense, RRF, and interpolation are scored once as a union, and each arm is then re-sorted by the cross-encoder score, so the arms differ only in their candidate source and share an identical reranking budget (all four reranked arms are listed in Supplementary Table~S10). Quora is subsampled to 1{,}000 queries (seed 42) for cost. The reranker runs in 16-bit floating point; we verified on samples that its rankings agree with 32-bit scoring to within 0.004 nDCG@10. Efficiency is measured on a single consumer workstation (Apple M5, 16\,GB unified memory, no discrete GPU): retrieval and fusion latencies on its CPU cores (batch size one, warm caches; median and 95th percentile over 200 queries), and reranker throughput on its integrated GPU; index sizes are measured on disk (Lucene) or as the embedding matrix (dense).

\subsection{Evaluation Metrics and Statistics}
nDCG@10, the standard BEIR cutoff, is the primary metric throughout, computed with \texttt{pytrec\_eval}~\cite{vangysel2018pytreceval}; recall@10 is reported where coverage matters. Pairwise comparisons use Wilcoxon signed-rank tests with Bonferroni correction and paired Cohen's $d$; 95\% confidence intervals use 1{,}000 bootstrap resamples of queries; risk-sensitive comparisons against the deployable baseline use URisk/TRisk~\cite{wang2012robust,dincer2014trisk}; variance decomposition uses one-way ANOVA $\eta^2$.

\subsection{Query Features for Predictability}
\label{sec:features}
We extract 17 features per query in three groups. \textbf{Query-intrinsic} (6): query length, IDF mean/max/std, entity density, distance to the corpus centroid. \textbf{Retrieval score-based} (5): BM25 and dense top-1 minus top-2 margins, mean and standard deviation of the dense top-10 similarities, top-10 Jaccard overlap between the two rankings. \textbf{QPP-inspired} (6): NQC and WIG~\cite{shtok2012qpp,zhou2007wig} for each retriever, their ratio, and a margin-based score-ratio weight. A gradient-boosted tree ensemble (GBT; 200 trees, depth 3) predicts fusion failure, defined as $\tau(q) > 0.01$ to exclude near-ties, and is evaluated by held-out ROC-AUC both within a corpus (5-fold) and across corpora (leave-one-corpus-out).

\section{Results}
\label{sec:results}

\subsection{Baseline Validation}
\label{sec:baseline-validation}
Table~\ref{tab:baselines} validates the single retrievers. Our BM25 runs reproduce the published Anserini BM25 nDCG@10~\cite{kamalloo2024brewingbeir} to four decimals on all seven corpora, so the sparse side is exactly the reference pipeline used throughout the BEIR literature. The Python \texttt{rank\_bm25} runs are 0.01--0.20 lower, most severely on TREC-COVID (0.459 vs.\ 0.656); Supplementary Table~S1 repeats the headline per-query analysis under all three BM25 implementations and shows that every qualitative finding below survives the swap, with one substantive change that we flag where it occurs (TREC-COVID, Section~\ref{sec:fusion-results}). Dense nDCG@10 rises from BGE-small to BGE-base on six corpora and from base to large on FiQA (0.391$\to$0.443), and is otherwise flat within 0.005 or slightly lower (by up to 0.028 on TREC-COVID); the exception is Quora, where the prefix-free protocol costs BGE-base and BGE-large 0.06 that the query instruction restores (Section~\ref{sec:ladder}, Supplementary Table~S8). The 2025-generation Qwen3-Embedding, run with its model card's generic retrieval instruction, is the strongest encoder in the study on FiQA, ArguAna, SciDocs, and TREC-COVID (0.467, 0.674, 0.227, 0.892) and weaker than the BGE models on SciFact, NFCorpus, and Quora (0.699, 0.355, 0.872). The ladder therefore provides genuinely stronger retrieval on most corpora, with diminishing returns within the BGE family.

\begin{table*}[!t]
\caption{Baseline validation: single-retriever nDCG@10 on all seven datasets. Anserini BM25 (ours) vs.\ the published Anserini regression values~\cite{kamalloo2024brewingbeir}; the \texttt{rank\_bm25} runs used for the implementation-sensitivity check; and dense retrieval across the five encoders.}
\label{tab:baselines}
\centering
{\footnotesize\setlength{\tabcolsep}{4pt}%
\begin{tabular}{lccc ccccc}
\toprule
 & \multicolumn{3}{c}{BM25 (nDCG@10)} & \multicolumn{5}{c}{Dense (nDCG@10)} \\
\cmidrule(lr){2-4}\cmidrule(lr){5-9}
Dataset & Anserini (ours) & Anserini (publ.) & \texttt{rank\_bm25} & MiniLM & BGE-sm & BGE-base & BGE-lg & Qwen3 \\
\midrule
SciFact & 0.6647 & 0.6647 & 0.6295 & 0.6451 & 0.7200 & 0.7376 & 0.7346 & 0.6992 \\
FiQA & 0.2361 & 0.2361 & 0.2016 & 0.3687 & 0.3848 & 0.3909 & 0.4432 & 0.4667 \\
NFCorpus & 0.3254 & 0.3254 & 0.2881 & 0.3159 & 0.3371 & 0.3681 & 0.3688 & 0.3552 \\
ArguAna & 0.4142 & 0.4142 & 0.3561 & 0.5017 & 0.5950 & 0.6362 & 0.6357 & 0.6737 \\
SciDocs & 0.1581 & 0.1581 & 0.1438 & 0.2164 & 0.1973 & 0.2045 & 0.2001 & 0.2274 \\
TREC-COVID & 0.6559 & 0.6559 & 0.4593 & 0.4725 & 0.6660 & 0.6719 & 0.6443 & 0.8923 \\
Quora & 0.7886 & 0.7886 & 0.7368 & 0.8755 & 0.8878 & 0.8280 & 0.8289 & 0.8717 \\
\bottomrule
\end{tabular}}
\end{table*}

\begin{table*}[!t]
\caption{Held-out selection of the interpolation weight (BGE-small). In-sample: best $\alpha$ on the evaluation queries; LOCO: $\alpha$ chosen on the other six corpora; CV: within-corpus 5-fold, mode of the selected weights. Tax vs.\ dense is the fraction of queries on which fusion scores below dense-only at the given weight; Helped is the fraction it improves at the LOCO weight.}
\label{tab:alpha}
\centering
{\footnotesize\setlength{\tabcolsep}{3.5pt}%
\begin{tabular}{l ccc cccc ccc c}
\toprule
 & \multicolumn{3}{c}{Selected $\alpha$} & \multicolumn{4}{c}{nDCG@10} & \multicolumn{3}{c}{Tax vs.\ dense (\%)} & Helped \\
\cmidrule(lr){2-4}\cmidrule(lr){5-8}\cmidrule(lr){9-11}\cmidrule(lr){12-12}
Dataset & In-sample & LOCO & CV & Dense & $\alpha{=}0.7$ & LOCO & CV & $\alpha{=}0.7$ & LOCO & CV & LOCO (\%) \\
\midrule
SciFact & 0.7 & 0.6 & 0.7 & 0.720 & 0.731 & 0.727 & 0.721 & 10.0 & 12.7 & 11.3 & 15.7 \\
FiQA & 0.7 & 0.6 & 0.7 & 0.385 & 0.409 & 0.405 & 0.409 & 14.7 & 20.7 & 14.7 & 23.9 \\
NFCorpus & 0.6 & 0.7 & 0.6 & 0.337 & 0.355 & 0.355 & 0.358 & 18.3 & 18.3 & 20.7 & 33.4 \\
ArguAna & 0.8 & 0.6 & 0.8 & 0.595 & 0.603 & 0.591 & 0.605 & 19.1 & 25.0 & 13.9 & 22.7 \\
SciDocs & 0.7 & 0.6 & 0.7 & 0.197 & 0.207 & 0.205 & 0.207 & 20.9 & 24.6 & 20.9 & 28.1 \\
TREC-COVID & 0.5 & 0.7 & 0.5 & 0.666 & 0.717 & 0.717 & 0.748 & 14.0 & 14.0 & 26.0 & 72.0 \\
Quora & 0.7 & 0.6 & 0.7 & 0.888 & 0.900 & 0.896 & 0.900 & 8.2 & 10.6 & 8.2 & 12.8 \\
\bottomrule
\end{tabular}}
\end{table*}

\subsection{Fusion Method Comparison and the Interpolation Weight}
\label{sec:fusion-results}
Table~\ref{tab:fusion} reports nDCG@10 for every method with BGE-small. The aggregate picture replicates Bruch et al.~\cite{bruch2023fusion} where their conditions hold and departs from it where they do not. Dense-weighted interpolation ($\alpha=0.7$) is the best fusion method on five of the seven corpora, all of them corpora on which dense retrieval leads BM25 clearly, by 0.04--0.18 nDCG@10 (0.039 on SciDocs): SciFact, FiQA, ArguAna, SciDocs, and Quora; its advantage over RRF is significant on four of the five after Bonferroni correction (Supplementary Table~S4: FiQA, ArguAna, SciDocs, Quora, $p<0.001$; SciFact $p=0.07$). On the two corpora where the two components are balanced, the picture reverses: on NFCorpus (BM25 0.325 vs.\ dense 0.337) the three fusion methods are within 0.003 of one another, and on TREC-COVID (0.656 vs.\ 0.666) rank fusion and CombMNZ (0.801 and 0.805) beat dense-heavy interpolation (0.717) and dense-only (0.666) by a wide and significant margin ($p=0.002$). This is the classical prediction of the calibration literature (Section~\ref{sec:related-calibration}): score interpolation wins when one component is reliably stronger and its confidence should be kept, rank fusion when neither should dominate. RRF is also frequently no better than the dense retriever alone: it trails dense-only on five corpora (significantly on ArguAna and Quora) and beats it only on the two balanced ones.

\begin{table*}[!t]
\caption{nDCG@10 across retrieval and fusion methods (BGE-small, Anserini BM25). Best per row in bold.}
\label{tab:fusion}
\centering
\begin{tabular}{lccccccc}
\toprule
Dataset & BM25 & Dense & RRF & Interp(0.3) & Interp(0.5) & Interp(0.7) & CombMNZ \\
\midrule
SciFact & 0.6647 & 0.7200 & 0.7152 & 0.7113 & 0.7273 & \textbf{0.7307} & 0.7230 \\
FiQA & 0.2361 & 0.3848 & 0.3667 & 0.3157 & 0.3886 & \textbf{0.4087} & 0.3872 \\
NFCorpus & 0.3254 & 0.3371 & 0.3570 & 0.3493 & 0.3560 & 0.3547 & \textbf{0.3574} \\
ArguAna & 0.4142 & 0.5950 & 0.5549 & 0.5115 & 0.5717 & \textbf{0.6027} & 0.5702 \\
SciDocs & 0.1581 & 0.1973 & 0.1954 & 0.1839 & 0.1999 & \textbf{0.2070} & 0.1976 \\
TREC-COVID & 0.6559 & 0.6660 & 0.8010 & 0.7077 & 0.7475 & 0.7166 & \textbf{0.8045} \\
Quora & 0.7886 & 0.8878 & 0.8695 & 0.8561 & 0.8875 & \textbf{0.8997} & 0.8857 \\
\bottomrule
\end{tabular}

\end{table*}

\textbf{The interpolation optimum is interior and transfers across corpora.}
\label{sec:alpha}
Two questions arise: whether the optimum is really interior rather than dense-only, and whether a weight chosen on one set of queries survives on another. Sweeping $\alpha$ from 0.5 to 1.0 (Supplementary Table~S7) answers the first: the in-sample optimum is $0.7$ on four corpora, $0.6$ on NFCorpus, $0.8$ on ArguAna, and $0.5$ on TREC-COVID, never $1.0$, and dense-only is strictly worse than the best interpolation on every corpus (e.g.\ FiQA 0.409 vs.\ 0.385, ArguAna 0.605 vs.\ 0.595, TREC-COVID 0.748 vs.\ 0.666); per query the optimal weight is rarely unique, because nDCG@10 is discrete: on 23--58\% of queries every weight on the 0.1 grid scores the same (2\% on TREC-COVID), dense-only ($\alpha=1$) is among the optimal weights on 52--84\% of queries on six corpora (18\% on TREC-COVID), some fused weight strictly beats dense-only on 16--48\% (82\% on TREC-COVID), a strictly interior weight is required on 8--30\% (70\% on TREC-COVID), and a near-dense weight ($\alpha\ge0.9$) is strictly required on only 6--16\%. Table~\ref{tab:alpha} answers the second with two held-out protocols. Under leave-one-corpus-out (LOCO) selection, the weight chosen on the other six corpora is $0.6$ or $0.7$ for every held-out corpus, and applying it costs at most 0.014 nDCG@10 relative to the in-sample optimum on six corpora (the largest, ArguAna, 0.605 vs.\ 0.591) and 0.031 on TREC-COVID, the one corpus whose in-sample optimum ($\alpha=0.5$) lies outside the transferred range; under within-corpus 5-fold cross-validation the selected weight coincides with the in-sample optimum on every corpus (SciFact's folds vary between 0.5 and 0.9, with mode 0.7) and the held-out nDCG@10 is within 0.010 of it. At the held-out weights the deployable tax is 10.6--25.0\% (LOCO) and 8.2--26.0\% (CV), against 8.2--20.9\% at the fixed $\alpha=0.7$: on five corpora the transferred weight, being slightly less dense-heavy than the in-sample optimum, raises it by 2--6 points, and the oracle tax at the held-out weights is 14--48\% (LOCO) and 13--46\% (CV). A dense-heavy weight of 0.6--0.7 is therefore a defensible transferable default, and the tax reported below is not an artifact of selecting the weight on the evaluation queries; if anything, a weight chosen elsewhere pays slightly more. The one corpus that prefers a balanced weight, TREC-COVID, illustrates the trade-off: moving to $\alpha=0.5$ raises its mean nDCG@10 from 0.717 to 0.748 while raising the fraction of queries that fall below dense-only from 14\% to 26\%.

\begin{figure*}[!t]
    \centering
    \includegraphics[width=\textwidth]{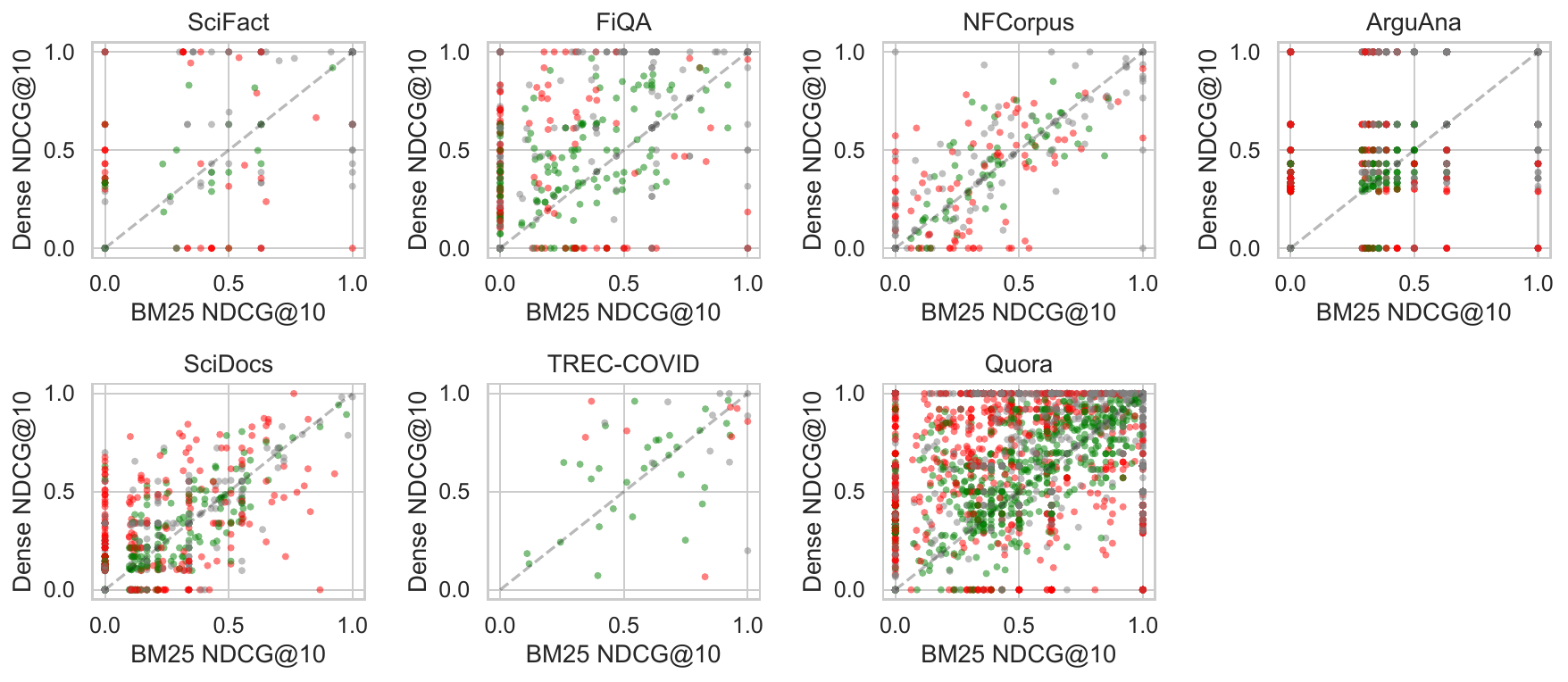}
    \caption{Per-query BM25 vs.\ dense nDCG@10 (BGE-small). For visualization only, each query is colored by the best of the three fusion methods with a 0.01 deadband: green, fusion helped ($\tau<-0.01$); red, fusion hurt ($\tau>0.01$); gray, neutral. Harm concentrates off the diagonal, where the retrievers disagree.}
    \label{fig:quadrant}
\end{figure*}

\subsection{Diagnostic Taxonomy}
\label{sec:taxonomy-results}
Table~\ref{tab:quadrants} gives the quadrant distribution and Fig.~\ref{fig:quadrant} the per-query landscape. \textbf{Both-Win} ranges from 42\% (FiQA) to 100\% (TREC-COVID). \textbf{Dense-Only} is substantial on the paraphrase-heavy and question-style corpora (ArguAna 20\%, FiQA 23\%, SciDocs 17\%), whereas \textbf{Sparse-Only} never exceeds 7\%: dense retrieval rarely misses a query that BM25 covers, even against a standard BM25. This asymmetry is itself a caution for hybridization: the queries covered by sparse alone form only a small share, although fusion can also recover coverage in Both-Fail (12 queries on ArguAna, 8 on SciDocs); its reordering upside lives in the much larger Both-Win quadrant, where a query can be helped or hurt, and its dilution risk lives in Both-Win too (on TREC-COVID, with no Sparse-Only queries at all, fusion still helps 36 of 50 queries against dense-only). \textbf{Both-Fail} tracks corpus difficulty (SciDocs 34\%, FiQA 30\%, NFCorpus 24\%). Fusion failures (red in Fig.~\ref{fig:quadrant}) concentrate off the diagonal, where one retriever succeeds and the other contributes noise.

\begin{table}[t]
\caption{Diagnostic quadrant distribution (\% of queries, BGE-small, Anserini BM25).}
\label{tab:quadrants}
\centering
{\footnotesize\setlength{\tabcolsep}{4pt}
\begin{tabular}{lcccc}
\toprule
Dataset & Both-Win & Dense-Only & Sparse-Only & Both-Fail \\
\midrule
SciFact & 76.3\% & 9.3\% & 5.3\% & 9.0\% \\
FiQA & 41.5\% & 22.5\% & 6.3\% & 29.6\% \\
NFCorpus & 62.5\% & 7.4\% & 6.5\% & 23.5\% \\
ArguAna & 65.3\% & 20.1\% & 5.0\% & 9.6\% \\
SciDocs & 42.8\% & 17.0\% & 6.7\% & 33.5\% \\
TREC-COVID & 100.0\% & 0.0\% & 0.0\% & 0.0\% \\
Quora & 92.2\% & 5.9\% & 1.0\% & 0.9\% \\
\bottomrule
\end{tabular}
}

\end{table}

\begin{figure*}[!t]
    \centering
    \includegraphics[width=\textwidth]{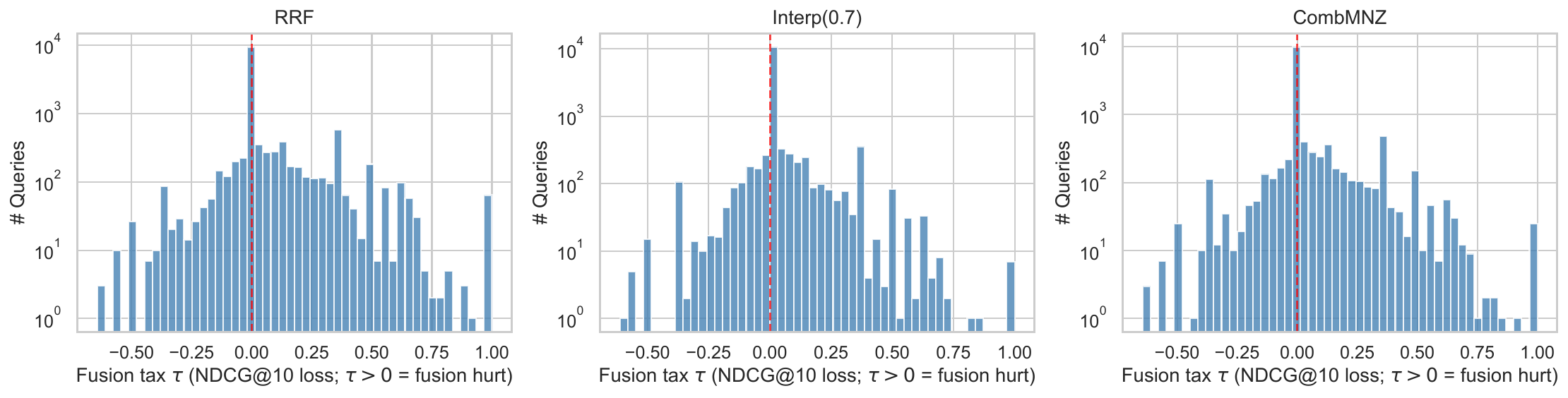}
    \caption{Distribution of the per-query fusion tax $\tau$ (Eq.~\ref{eq:tax}) pooled over the seven datasets (BGE-small). Positive values are queries that fusion hurt; the long right tail (log-scale counts) shows that when fusion hurts, the loss can exceed 0.5 nDCG@10.}
    \label{fig:tax-hist}
\end{figure*}

\begin{table*}[!t]
\caption{The interpolation tax by diagnostic quadrant (BGE-small, vs.\ the per-query oracle). Sh.: quadrant share of queries (\%); Hurt/Help: \% of the quadrant's queries hurt/helped; Net: mean $\Delta$nDCG@10 within the quadrant.}
\label{tab:quadrant-tax}
\centering
{\footnotesize\setlength{\tabcolsep}{2.6pt}%
\begin{tabular}{l cccc cccc cccc cccc}
\toprule
 & \multicolumn{4}{c}{Both-Win} & \multicolumn{4}{c}{Dense-Only} & \multicolumn{4}{c}{Sparse-Only} & \multicolumn{4}{c}{Both-Fail} \\
\cmidrule(lr){2-5}\cmidrule(lr){6-9}\cmidrule(lr){10-13}\cmidrule(lr){14-17}
Dataset & Sh. & Hurt & Help & Net & Sh. & Hurt & Help & Net & Sh. & Hurt & Help & Net & Sh. & Hurt & Help & Net \\
\midrule
SciFact & 76 & 12 & 4 & -0.024 & 9 & 61 & 7 & -0.107 & 5 & 94 & 0 & -0.319 & 9 & 0 & 0 & +0.000 \\
FiQA & 42 & 22 & 28 & +0.008 & 23 & 38 & 16 & -0.029 & 6 & 90 & 5 & -0.271 & 30 & 0 & 4 & +0.008 \\
NFCorpus & 63 & 45 & 25 & -0.026 & 7 & 62 & 8 & -0.035 & 7 & 95 & 5 & -0.220 & 24 & 0 & 0 & +0.000 \\
ArguAna & 65 & 23 & 10 & -0.031 & 20 & 56 & 12 & -0.096 & 5 & 87 & 7 & -0.243 & 10 & 0 & 9 & +0.027 \\
SciDocs & 43 & 44 & 24 & -0.020 & 17 & 65 & 16 & -0.041 & 7 & 94 & 2 & -0.142 & 34 & 0 & 2 & +0.003 \\
TREC-COVID & 100 & 48 & 40 & -0.041 & 0 & -- & -- & -- & 0 & -- & -- & -- & 0 & -- & -- & -- \\
Quora & 92 & 10 & 5 & -0.011 & 6 & 47 & 7 & -0.092 & 1 & 88 & 4 & -0.371 & 1 & 0 & 4 & +0.015 \\
\bottomrule
\end{tabular}}
\end{table*}

\begin{table*}[!t]
\caption{Symmetric harm/benefit accounting of fusion (BGE-small). Hurt/Help: percentage of queries with lower/higher nDCG@10 than the reference; Loss/Gain: mean absolute change on those queries; Net $\Delta$: mean change over all queries (Eq.~\ref{eq:net}).}
\label{tab:harm}
\centering
{\footnotesize\setlength{\tabcolsep}{3pt}%
\begin{tabular}{l ccccc ccccc}
\toprule
 & \multicolumn{5}{c}{vs.\ always-dense (deployable)} & \multicolumn{5}{c}{vs.\ per-query oracle $\max(\text{BM25},\text{dense})$} \\
\cmidrule(lr){2-6}\cmidrule(lr){7-11}
Dataset & Hurt\,\% & Loss & Help\,\% & Gain & Net $\Delta$ & Hurt\,\% & Loss & Help\,\% & Gain & Net $\Delta$ \\
\midrule
\multicolumn{11}{l}{\emph{Interp(0.7)}} \\
SciFact & 10.0 & 0.205 & 14.7 & 0.213 & +0.0107 & 20.0 & 0.250 & 3.7 & 0.128 & -0.0453 \\
FiQA & 14.7 & 0.100 & 23.1 & 0.166 & +0.0239 & 23.3 & 0.157 & 16.5 & 0.111 & -0.0181 \\
NFCorpus & 18.3 & 0.044 & 33.4 & 0.077 & +0.0177 & 39.0 & 0.103 & 16.4 & 0.042 & -0.0332 \\
ArguAna & 19.1 & 0.213 & 21.8 & 0.221 & +0.0077 & 30.4 & 0.221 & 10.3 & 0.172 & -0.0492 \\
SciDocs & 20.9 & 0.070 & 26.1 & 0.093 & +0.0097 & 36.2 & 0.090 & 14.0 & 0.060 & -0.0243 \\
TREC-COVID & 14.0 & 0.054 & 72.0 & 0.081 & +0.0506 & 48.0 & 0.140 & 40.0 & 0.066 & -0.0407 \\
Quora & 8.2 & 0.167 & 12.4 & 0.205 & +0.0119 & 12.8 & 0.206 & 5.5 & 0.131 & -0.0192 \\
\midrule
\multicolumn{11}{l}{\emph{RRF}} \\
SciFact & 15.3 & 0.357 & 18.0 & 0.278 & -0.0048 & 23.3 & 0.311 & 6.3 & 0.187 & -0.0608 \\
FiQA & 30.4 & 0.269 & 26.9 & 0.237 & -0.0181 & 35.8 & 0.262 & 20.5 & 0.165 & -0.0601 \\
NFCorpus & 23.8 & 0.110 & 37.8 & 0.122 & +0.0199 & 39.0 & 0.106 & 16.7 & 0.063 & -0.0309 \\
ArguAna & 32.9 & 0.316 & 23.7 & 0.269 & -0.0401 & 40.1 & 0.299 & 11.7 & 0.198 & -0.0970 \\
SciDocs & 27.6 & 0.139 & 30.1 & 0.121 & -0.0018 & 38.1 & 0.133 & 18.8 & 0.078 & -0.0358 \\
TREC-COVID & 26.0 & 0.096 & 70.0 & 0.228 & +0.1350 & 36.0 & 0.087 & 54.0 & 0.139 & +0.0437 \\
Quora & 16.2 & 0.293 & 12.5 & 0.234 & -0.0184 & 19.9 & 0.288 & 5.4 & 0.142 & -0.0494 \\
\bottomrule
\end{tabular}}
\end{table*}

\subsection{The Fusion Tax: Harm and Benefit}
\label{sec:fusion-tax-results}
Table~\ref{tab:harm} gives the symmetric accounting of Section~\ref{sec:fusion-tax} for interpolation and RRF against both references, and Fig.~\ref{fig:tax-hist} the distribution of $\tau$.

\textbf{Oracle tax.} Against the per-query oracle, the best method (interpolation, $\alpha=0.7$) hurts 12.8\% of queries on the easy, dense-dominated Quora set, 20--39\% on five corpora, and 48\% on the 50-query TREC-COVID set; RRF hurts 20--40\% and CombMNZ 17--40\% (per-method rates and mean losses in Supplementary Table~S9). These are not near-ties: requiring a material loss of $\tau>0.05$ nDCG@10 still leaves 10--34\% of queries (interpolation) and $\tau>0.10$ leaves 8--22\% (Supplementary Table~S9), and the mean loss on a hurt query is 0.09--0.25 nDCG@10. Bootstrap confidence intervals (Supplementary Table~S2) sit well above zero on every corpus. Fusion beats \emph{both} of its components on only 4--40\% of queries, so the net change against the oracle is negative everywhere ($-0.018$ to $-0.049$): a perfect per-query router would gain 0.02--0.05 nDCG@10 over static fusion on every corpus, which is the prize that motivates routing and that Section~\ref{sec:feature-results} shows is hard to collect.

\textbf{Deployable tax and benefit.} Against always-dense, the reference a practitioner can actually deploy, interpolation hurts 8--21\% of queries and helps 12--72\%. The mean gain on a helped query equals or exceeds the mean loss on a hurt query on every corpus, by under 5\% on SciFact and ArguAna (0.213 vs.\ 0.205, 0.221 vs.\ 0.213) and by 1.2--1.8$\times$ on the others (e.g.\ FiQA 0.166 vs.\ 0.100, NFCorpus 0.077 vs.\ 0.044), so fusion's aggregate edge, positive on all seven corpora ($+0.008$ to $+0.051$), comes from helping somewhat more queries, and somewhat more per query, than it hurts. The losses are nevertheless of the same order as the gains, 0.04--0.21 nDCG@10 on a hurt query, not small change. For RRF the balance tips: it hurts 15--33\% of queries and its net change against dense-only is negative on five corpora ($-0.002$ to $-0.039$), positive only on the two balanced corpora (NFCorpus $+0.020$, TREC-COVID $+0.135$).

\textbf{By quadrant.} Table~\ref{tab:quadrant-tax} decomposes the interpolation tax by quadrant. In \textbf{Sparse-Only}, dense-heavy fusion hurts 87--95\% of queries with a net loss of 0.14--0.37 nDCG@10: the dense side contributes only noise there and a weight of 0.7 buries BM25's relevant documents. In \textbf{Dense-Only} it hurts 38--65\% (net $-0.03$ to $-0.11$), the mirror image with BM25 as the noise source. In \textbf{Both-Win} the two directions nearly cancel (net $-0.04$ to $+0.01$) while 10--48\% of queries are still reordered for the worse, and in \textbf{Both-Fail} fusion can only help, which it does on 0--9\% of queries. The harm \emph{rate} is thus highest exactly where Fig.~\ref{fig:quadrant} places it: off the diagonal, on the at most 29\% of queries that only one retriever covers, and the dense-heavy weight that wins in aggregate does so by sacrificing the Sparse-Only minority. Rate and count differ, however: because Both-Win is by far the largest quadrant, it holds 38--72\% of the hurt queries and 26--63\% of the total loss (all of both on TREC-COVID, whose 50 queries are all Both-Win), so fixing only the off-diagonal queries would leave most of the tax in place (Supplementary Table~S14).

\textbf{Mostly ranking, not coverage.} Fusion's recall@10 equals or exceeds dense-only's in aggregate on all seven corpora (Supplementary Table~S5), and recall is reduced on only 1--7\% of all queries. The hurt queries themselves are less clean: among the queries interpolation hurts relative to dense-only, 10--43\% also lose at least one relevant document from the top-10, those queries carry 15--59\% of the total nDCG@10 downside (15--35\% on four corpora, 48--59\% on NFCorpus, SciDocs, and TREC-COVID), and 77 queries across the seven corpora lose every relevant document from the top-10 (Supplementary Table~S5). For most hurt queries the relevant documents remain in the top-10 and fusion reorders them, so the harm falls mainly on consumers sensitive to rank order, but a minority of it is a coverage loss that no downstream reordering can repair; Section~\ref{sec:rerank-results} measures how much of the rest a second stage removes.

\textbf{Risk-sensitive significance.} When the deployable comparison is placed on the standard risk-sensitive footing~\cite{wang2012robust,dincer2014trisk}, TRisk (Supplementary Table~S3) flags interpolation as significantly risky relative to always-dense at risk aversion $\alpha_{\text{risk}}=5$ on six of the seven corpora (all but TREC-COVID) even though it beats dense-only on the mean, and already at $\alpha_{\text{risk}}=1$ on ArguAna; RRF is significantly risky on five corpora at $\alpha_{\text{risk}}=1$ and six at $\alpha_{\text{risk}}=5$. TREC-COVID, where both components are strong, is the one corpus on which both fusions are risk-robust.

\begin{figure*}[!t]
    \centering
    \includegraphics[width=0.92\linewidth]{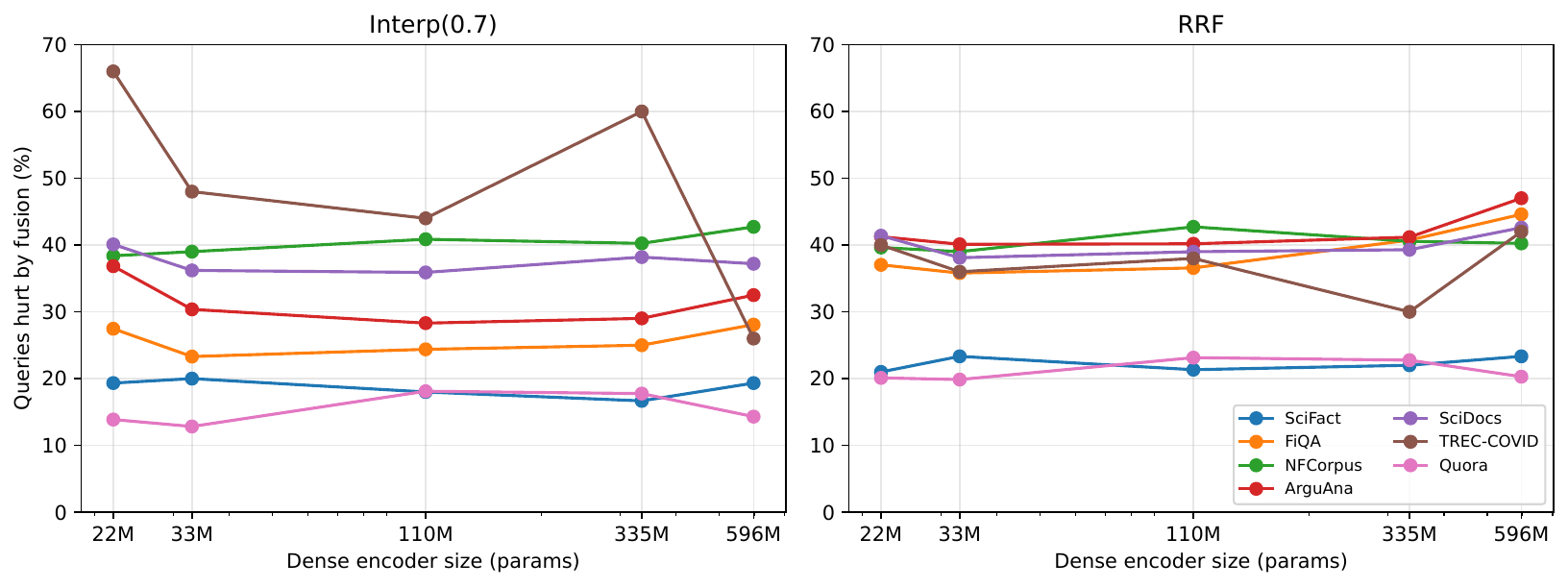}
    \caption{Fusion tax (\% of queries hurt) as a function of dense-encoder size (log scale) for interpolation (left) and RRF (right), all seven corpora. The tax does not shrink as encoders scale from 22M to 596M parameters.}
    \label{fig:ladder}
\end{figure*}

\subsection{Robustness Across the Encoder Scale Ladder}
\label{sec:ladder}
A natural objection is that the tax reflects weak dense encoders. We scale the dense encoder by an order of magnitude within one family (BGE-small 33M, base 110M, large 335M), add MiniLM (22M) as a cross-family check and Qwen3-Embedding (596M, released 2025) as a newer-generation check, and recompute the tax on all seven corpora (Table~\ref{tab:ladder}, Fig.~\ref{fig:ladder}).

The tax does not vanish. For interpolation it is flat within the BGE family, drifting down on SciFact and ArguAna and up on FiQA and NFCorpus by at most four points (small, base, large: SciFact 20.0, 18.0, 16.7\%; FiQA 23.3, 24.4, 25.0\%; NFCorpus 39.0, 40.9, 40.2\%; ArguAna 30.4, 28.3, 29.0\%), and for RRF it is flat to rising (FiQA 35.8, 36.6, 40.7\%), even though the larger encoders deliver substantially higher absolute nDCG@10. The other three corpora show the same pattern: for interpolation, SciDocs 36.2, 35.9, 38.2\%, Quora 12.8, 18.1, 17.8\%, and the 50-query TREC-COVID set 48, 44, 60\%, with RRF at 38--39\% on SciDocs, 30--38\% on TREC-COVID, and 20--23\% on Quora. At BGE-large, the strongest BGE rung, interpolation still hurts 17--40\% of queries on six corpora (60\% on TREC-COVID) and RRF 22--41\%, while fusion continues to beat dense-only in aggregate on every corpus. One protocol caveat concerns Quora: encoded prefix-free, BGE-base and BGE-large score 0.83 there against 0.89 for BGE-small. Re-encoding only the queries with the BGE instruction restores 0.887--0.890 and brings their Quora tax to 13\%, the same as BGE-small's; the instruction also raises the TREC-COVID dense scores of all three BGE models by 0.09--0.11 and moves the tax by at most three points on the other five corpora (Supplementary Table~S8), so the conclusion does not depend on the prefix protocol. Qwen3-Embedding, a 596M-parameter model of the 2025 generation, continues the pattern rather than breaking it: interpolation hurts 14--43\% of its queries (SciFact 19\%, FiQA 28\%, NFCorpus 43\%, ArguAna 33\%, SciDocs 37\%, TREC-COVID 26\%, Quora 14\%) and RRF 20--47\%, while fusion beats Qwen3 alone in aggregate on five corpora, falls marginally short of it on SciDocs (0.225 vs.\ 0.227), and loses to it on ArguAna (0.638 vs.\ 0.674), where Qwen3 is the strongest encoder in the study and even the dense-heavy weight dilutes it with a BM25 that scores 0.414. Where its tax differs from BGE-small's, it moves with the sparse--dense balance rather than with encoder quality: it is highest where Qwen3 barely leads BM25 (NFCorpus, 0.36 vs.\ 0.33; 43\%) and lowest where Qwen3 dominates (TREC-COVID, 0.89 vs.\ 0.66, 26\%; Quora, 0.87 vs.\ 0.79, 14\%), exactly as Section~\ref{sec:fusion-results} predicts; ArguAna, where Qwen3 leads BM25 by 0.26, sits in between at 33\% and has the largest aggregate loss from interpolation relative to Qwen3; SciDocs also shows a smaller loss (Table~\ref{tab:ladder}). The five-encoder picture is consistent on the claim that matters: strengthening the dense retriever by an order of magnitude, or moving to a newer generation, never removes the tax, and what changes it is the balance between the two retrievers. On ArguAna, where a weak BM25 had let dense-only beat fusion at every scale, the standard BM25 closes that gap (dense 0.636 vs.\ interpolation 0.636--0.638 at base and large), yet 28--29\% of its queries still pay the tax.

\begin{table*}[!t]
\caption{The fusion tax across five encoders spanning 22M--596M parameters and three model generations, all seven corpora. BGE-small/base/large form a controlled same-family scale ladder; MiniLM and Qwen3-Embedding are cross-family and cross-generation checks. ``Fusion tax'' is the oracle tax: the percentage of queries each fusion method hurts relative to the better component.}
\label{tab:ladder}
\centering
\begin{tabular}{llccc cc}
\toprule
 & & \multicolumn{3}{c}{nDCG@10} & \multicolumn{2}{c}{Fusion tax (\% hurt)} \\
\cmidrule(lr){3-5}\cmidrule(lr){6-7}
Dataset & Encoder & Dense & Interp(0.7) & RRF & Interp(0.7) & RRF \\
\midrule
\multirow{5}{*}{SciFact} & MiniLM (22M) & 0.645 & 0.693 & 0.700 & 19.3\% & 21.0\% \\
 & BGE-small (33M) & 0.720 & 0.731 & 0.715 & 20.0\% & 23.3\% \\
 & BGE-base (110M) & 0.738 & 0.757 & 0.731 & 18.0\% & 21.3\% \\
 & BGE-large (335M) & 0.735 & 0.755 & 0.736 & 16.7\% & 22.0\% \\
 & Qwen3-Emb (596M, 2025) & 0.699 & 0.728 & 0.716 & 19.3\% & 23.3\% \\
\midrule
\multirow{5}{*}{FiQA} & MiniLM (22M) & 0.369 & 0.400 & 0.369 & 27.5\% & 37.0\% \\
 & BGE-small (33M) & 0.385 & 0.409 & 0.367 & 23.3\% & 35.8\% \\
 & BGE-base (110M) & 0.391 & 0.416 & 0.374 & 24.4\% & 36.6\% \\
 & BGE-large (335M) & 0.443 & 0.459 & 0.399 & 25.0\% & 40.7\% \\
 & Qwen3-Emb (596M, 2025) & 0.467 & 0.469 & 0.389 & 28.1\% & 44.6\% \\
\midrule
\multirow{5}{*}{NFCorpus} & MiniLM (22M) & 0.316 & 0.335 & 0.350 & 38.4\% & 39.6\% \\
 & BGE-small (33M) & 0.337 & 0.355 & 0.357 & 39.0\% & 39.0\% \\
 & BGE-base (110M) & 0.368 & 0.378 & 0.370 & 40.9\% & 42.7\% \\
 & BGE-large (335M) & 0.369 & 0.380 & 0.371 & 40.2\% & 40.6\% \\
 & Qwen3-Emb (596M, 2025) & 0.355 & 0.364 & 0.360 & 42.7\% & 40.2\% \\
\midrule
\multirow{5}{*}{ArguAna} & MiniLM (22M) & 0.502 & 0.517 & 0.502 & 36.8\% & 41.3\% \\
 & BGE-small (33M) & 0.595 & 0.603 & 0.555 & 30.4\% & 40.1\% \\
 & BGE-base (110M) & 0.636 & 0.636 & 0.575 & 28.3\% & 40.2\% \\
 & BGE-large (335M) & 0.636 & 0.638 & 0.578 & 29.0\% & 41.2\% \\
 & Qwen3-Emb (596M, 2025) & 0.674 & 0.638 & 0.560 & 32.5\% & 47.0\% \\
\midrule
\multirow{5}{*}{SciDocs} & MiniLM (22M) & 0.216 & 0.220 & 0.202 & 40.1\% & 41.4\% \\
 & BGE-small (33M) & 0.197 & 0.207 & 0.195 & 36.2\% & 38.1\% \\
 & BGE-base (110M) & 0.205 & 0.214 & 0.199 & 35.9\% & 39.0\% \\
 & BGE-large (335M) & 0.200 & 0.218 & 0.207 & 38.2\% & 39.3\% \\
 & Qwen3-Emb (596M, 2025) & 0.227 & 0.225 & 0.201 & 37.2\% & 42.6\% \\
\midrule
\multirow{5}{*}{TREC-COVID} & MiniLM (22M) & 0.472 & 0.530 & 0.726 & 66.0\% & 40.0\% \\
 & BGE-small (33M) & 0.666 & 0.717 & 0.801 & 48.0\% & 36.0\% \\
 & BGE-base (110M) & 0.672 & 0.730 & 0.814 & 44.0\% & 38.0\% \\
 & BGE-large (335M) & 0.644 & 0.693 & 0.788 & 60.0\% & 30.0\% \\
 & Qwen3-Emb (596M, 2025) & 0.892 & 0.897 & 0.873 & 26.0\% & 42.0\% \\
\midrule
\multirow{5}{*}{Quora} & MiniLM (22M) & 0.875 & 0.893 & 0.867 & 13.9\% & 20.1\% \\
 & BGE-small (33M) & 0.888 & 0.900 & 0.869 & 12.8\% & 19.9\% \\
 & BGE-base (110M) & 0.828 & 0.842 & 0.850 & 18.1\% & 23.1\% \\
 & BGE-large (335M) & 0.829 & 0.842 & 0.851 & 17.8\% & 22.8\% \\
 & Qwen3-Emb (596M, 2025) & 0.872 & 0.889 & 0.865 & 14.3\% & 20.3\% \\
\bottomrule
\end{tabular}
\end{table*}

\subsection{Robustness to the Sparse Method: SPLADE}
\label{sec:splade}
The ladder rules out a weak dense model; we now rule out a weak sparse one by replacing BM25 with SPLADE~\cite{formal2022splade} as the sparse side of fusion with BGE-small on all seven corpora (Table~\ref{tab:splade}). Pairing a learned-sparse retriever with a dense one is not new (Bruch et al.\ include SPLADE among their components); the contribution is the per-component \emph{attribution} of the tax. SPLADE is a stronger sparse retriever than BM25 on six corpora (e.g.\ 0.708 vs.\ 0.665 on SciFact, 0.356 vs.\ 0.236 on FiQA, 0.725 vs.\ 0.656 on TREC-COVID) and ties it on SciDocs, and SPLADE+dense interpolation beats dense-only in aggregate on six corpora and ties it on Quora. Yet interpolation's tax does not abate: SPLADE+dense interpolation hurts 14--41\% of queries on six corpora and 60\% on TREC-COVID, comparable to or higher than BM25+dense on five of the seven and lower by under three points on the other two (ArguAna, SciDocs). \textbf{Interpolation's tax is thus intrinsic to score fusion}, robust to strengthening either side. For RRF the picture is different in kind. With the standard Anserini BM25 the sparse upgrade moves RRF's tax by only $-6.2$ to $+7.1$ points, lowering it on four corpora (most on ArguAna, 33.9\% vs.\ 40.1\%) and raising it on three. The large RRF reductions that upgrading a weak lexical baseline produces are visible in the BM25 swap itself: moving from \texttt{rank\_bm25} to Anserini lowered RRF's tax from 25--48\% to 20--40\% and raised its nDCG@10 on every corpus while changing interpolation's tax by at most 5.3 points (Supplementary Table~S1). \textbf{RRF's tax therefore responds to the quality of the lexical component} in a way interpolation's does not, which is consistent with rank fusion's inability to discount a weak component's confident-looking ranks; once the lexical side is standard, however, a substantial RRF tax remains (19--46\% with SPLADE).

\begin{table*}[!t]
\caption{Robustness to the sparse method (BGE-small). SPLADE replaces BM25 as the sparse side; ``Tax \%'' is the oracle tax; SP+De = SPLADE+dense, BM+De = BM25+dense.}
\label{tab:splade}
\centering
\begin{tabular}{l cccc cc cc}
\toprule
 & \multicolumn{4}{c}{nDCG@10} & \multicolumn{2}{c}{Tax \% (Interp 0.7)} & \multicolumn{2}{c}{Tax \% (RRF)} \\
\cmidrule(lr){2-5}\cmidrule(lr){6-7}\cmidrule(lr){8-9}
Dataset & BM25 & SPLADE & Dense & SP+De & SP+De & BM+De & SP+De & BM+De \\
\midrule
SciFact & 0.665 & 0.708 & 0.720 & 0.737 & 21.7\% & 20.0\% & 22.7\% & 23.3\% \\
FiQA & 0.236 & 0.356 & 0.385 & 0.411 & 29.6\% & 23.3\% & 32.7\% & 35.8\% \\
NFCorpus & 0.325 & 0.352 & 0.337 & 0.362 & 41.2\% & 39.0\% & 46.1\% & 39.0\% \\
ArguAna & 0.414 & 0.508 & 0.595 & 0.619 & 27.8\% & 30.4\% & 33.9\% & 40.1\% \\
SciDocs & 0.158 & 0.158 & 0.197 & 0.203 & 35.7\% & 36.2\% & 39.4\% & 38.1\% \\
TREC-COVID & 0.656 & 0.725 & 0.666 & 0.739 & 60.0\% & 48.0\% & 40.0\% & 36.0\% \\
Quora & 0.789 & 0.834 & 0.888 & 0.888 & 13.5\% & 12.8\% & 18.8\% & 19.9\% \\
\bottomrule
\end{tabular}
\end{table*}

\begin{table*}[!t]
\caption{Per-corpus GBT feature importance for predicting interpolation failure (bold: top-3 per corpus). Superscripts give the sign of the feature's Spearman association with failure ($|\rho|\ge0.05$); the last column counts corpora with positive/negative association.}
\label{tab:feature-importance}
\centering
{\footnotesize\setlength{\tabcolsep}{3pt}%
\begin{tabular}{l ccccccc c}
\toprule
Feature & SciFact & FiQA & NFCorpus & ArguAna & SciDocs & TREC-COVID & Quora & $+$/$-$ \\
\midrule
dense margin & \textbf{0.18}$^{-}$ & 0.07$^{-}$ & 0.04 & \textbf{0.08}$^{-}$ & 0.06 & \textbf{0.14}$^{+}$ & \textbf{0.35}$^{-}$ & 1/4 \\
score-ratio $\alpha$ & 0.06$^{-}$ & \textbf{0.11} & 0.06$^{+}$ & 0.07$^{-}$ & 0.07 & 0.10$^{+}$ & 0.05$^{-}$ & 2/3 \\
mean IDF & \textbf{0.11}$^{-}$ & 0.05$^{+}$ & 0.08$^{-}$ & 0.07 & 0.07 & 0.09 & 0.03$^{-}$ & 1/3 \\
WIG (dense) & 0.04$^{+}$ & 0.06$^{+}$ & 0.07 & \textbf{0.08}$^{+}$ & \textbf{0.09}$^{+}$ & \textbf{0.10}$^{-}$ & 0.06$^{+}$ & 5/1 \\
WIG (BM25) & 0.09 & \textbf{0.09}$^{+}$ & 0.05$^{+}$ & \textbf{0.09}$^{+}$ & \textbf{0.10}$^{+}$ & 0.06$^{+}$ & 0.02 & 5/0 \\
centroid dist. & 0.03 & 0.07 & \textbf{0.12}$^{-}$ & 0.07 & 0.07$^{+}$ & 0.06$^{+}$ & 0.04 & 2/1 \\
dense top-10 mean & 0.06$^{+}$ & 0.07$^{+}$ & 0.07$^{+}$ & 0.07$^{+}$ & \textbf{0.09}$^{+}$ & 0.02 & 0.08$^{+}$ & 6/0 \\
IDF std & 0.05$^{-}$ & 0.06 & 0.05 & 0.06 & 0.06 & \textbf{0.13}$^{-}$ & 0.03$^{-}$ & 0/3 \\
BM25 margin & \textbf{0.11}$^{-}$ & 0.07$^{-}$ & 0.07$^{+}$ & 0.05 & 0.07 & 0.03$^{-}$ & 0.03$^{-}$ & 1/4 \\
QPP ratio & 0.05 & \textbf{0.08}$^{-}$ & 0.04$^{-}$ & 0.07$^{-}$ & 0.06 & 0.03$^{-}$ & 0.03 & 0/4 \\
dense top-10 std & 0.04$^{-}$ & 0.03 & 0.05 & 0.04$^{-}$ & 0.05$^{+}$ & 0.05$^{-}$ & \textbf{0.09}$^{-}$ & 1/4 \\
NQC (BM25) & 0.04$^{-}$ & 0.06$^{+}$ & \textbf{0.08}$^{+}$ & 0.06$^{+}$ & 0.05$^{+}$ & 0.01$^{+}$ & 0.03$^{-}$ & 5/2 \\
NQC (dense) & 0.05$^{-}$ & 0.03 & 0.05$^{-}$ & 0.05$^{-}$ & 0.04$^{+}$ & 0.03$^{-}$ & \textbf{0.08}$^{-}$ & 1/5 \\
top-10 overlap & 0.02$^{+}$ & 0.03$^{+}$ & \textbf{0.08}$^{+}$ & 0.02$^{+}$ & 0.02$^{+}$ & 0.06$^{-}$ & 0.03$^{+}$ & 6/1 \\
entity density & 0.03 & 0.04$^{+}$ & 0.02 & 0.07 & 0.05 & 0.02 & 0.02 & 1/0 \\
query length & 0.04 & 0.04 & 0.01$^{+}$ & 0.06 & 0.02 & 0.02$^{-}$ & 0.01 & 1/1 \\
max IDF & 0.01$^{-}$ & 0.03 & 0.06$^{-}$ & 0.00 & 0.03 & 0.06$^{-}$ & 0.03$^{-}$ & 0/4 \\
\bottomrule
\end{tabular}}
\end{table*}

\subsection{Is Fusion Failure Predictable?}
\label{sec:feature-results}
If fusion failure were predictable from cheap query features, a practitioner could route around it. Table~\ref{tab:feature-auc} reports the held-out ROC-AUC of the gradient-boosted classifier of Section~\ref{sec:features}, within a corpus (5-fold) and across corpora (leave-one-corpus-out). \textbf{Within a corpus, failure is weakly predictable}: mean AUC 0.63 for the interpolation label and 0.58 for RRF, reaching 0.80 on Quora, so the features carry signal. \textbf{Across corpora, transfer is unreliable}: the mean drops to 0.59 and 0.57, but the mean hides the problem, which is variance. Cross-corpus AUC ranges from 0.49 (NFCorpus, chance) to 0.74 (SciFact) for interpolation and from 0.50 (NFCorpus) to 0.71 (SciFact) for RRF; it is at chance on NFCorpus for both labels and within 0.03 of chance on TREC-COVID. Capping every training corpus at 2{,}000 queries, so that the 10{,}000-query Quora set cannot dominate the pooled trainer, changes little (cross-corpus means 0.56 and 0.54), and replacing the 17 features by the 384-dimensional dense query embedding is at chance both within (0.53) and across (0.51) corpora (Supplementary Table~S6). The signal a router could use is therefore present on some targets and absent on others, nothing in the cheap features tells the practitioner which case they are in, and even where it is present the router tested next does not turn it into a gain.

The AUCs above concern the oracle-failure label. The deployable decision is different: whether switching a query to dense-only beats fusion. That label is predictable to a similar degree (mean AUC 0.61 within a corpus and 0.57 across, ranging from 0.33 on NFCorpus to 0.66 on Quora; Supplementary Table~S6), and a router that acts on it does not pay. Trained on the six other corpora with its threshold chosen on out-of-fold predictions of the training data, it never beats static interpolation on the held-out corpus: on three corpora (FiQA, SciDocs, TREC-COVID) no held-out query's predicted probability reaches the selected threshold, so it routes nothing, on two it routes 1--6\% of queries with no measurable effect, and on NFCorpus and Quora it routes 35\% and 23\% of queries and loses 0.007 and 0.003 nDCG@10 (paired 95\% intervals excluding zero). Within a corpus it stays within 0.001 of static interpolation everywhere, against an oracle gain of 0.007--0.041 (Supplementary Table~S11). This is a negative result for one classifier and threshold procedure on these features; it does not rule out a better router, but it does mean the tax cannot be routed around with the cheap signals we tested.

Table~\ref{tab:feature-importance} shows why. The features that matter differ by corpus: the dense top-1 margin dominates on SciFact, TREC-COVID, and Quora, the BM25 WIG predictor on ArguAna and SciDocs (and ranks second on FiQA), the score-ratio $\alpha$ on FiQA, and centroid distance on NFCorpus; the mean pairwise Spearman correlation between corpora's importance vectors is only 0.28. More damaging still, the \emph{direction} of association flips: 11 of the 17 features are positively associated with failure on at least one corpus and negatively on another (mean pairwise Spearman correlation of the association vectors 0.33). The dense margin is the clearest case: a confident dense top-1 predicts a \emph{safe} query on the dense-dominated corpora, where fusion can only dilute it, and a \emph{failing} query on TREC-COVID, where BM25 is equally strong and the dense-heavy weight discards evidence. The same signal has opposite meanings under different sparse--dense balances, which is exactly what a corpus-agnostic router cannot learn.

\begin{table}[t]
\caption{Predicting fusion failure from 17 query features (held-out ROC-AUC, BGE-small). In-domain: 5-fold within the corpus; cross-corpus: trained on the other six.}
\label{tab:feature-auc}
\centering
{\footnotesize\setlength{\tabcolsep}{4pt}%
\begin{tabular}{lcc cc}
\toprule
 & \multicolumn{2}{c}{Interp(0.7) failure} & \multicolumn{2}{c}{RRF failure} \\
\cmidrule(lr){2-3}\cmidrule(lr){4-5}
Dataset & In-domain & Cross-corpus & In-domain & Cross-corpus \\
\midrule
SciFact & 0.70 & 0.74 & 0.67 & 0.71 \\
FiQA & 0.58 & 0.55 & 0.55 & 0.57 \\
NFCorpus & 0.61 & 0.49 & 0.60 & 0.50 \\
ArguAna & 0.59 & 0.60 & 0.55 & 0.55 \\
SciDocs & 0.56 & 0.56 & 0.58 & 0.55 \\
TREC-COVID & 0.55 & 0.51 & 0.35 & 0.53 \\
Quora & 0.80 & 0.70 & 0.78 & 0.56 \\
\midrule
\textbf{Mean} & 0.63 & 0.59 & 0.58 & 0.57 \\
\bottomrule
\end{tabular}}
\end{table}

\subsection{Does Reranking Pay the Tax?}
\label{sec:rerank-results}
Table~\ref{tab:rerank} reports the budget-matched reranking experiment of Section~\ref{sec:setup-rerank}. The answer is: partly, and only where the reranker is itself reliable. On the three corpora where the cross-encoder improves every retrieval arm (SciFact, FiQA, TREC-COVID; dense-only rises by 0.019--0.142), reranking lowers the prevalence of interpolation's deployable tax by a third on SciFact and not at all on FiQA and TREC-COVID, where 16\% and 22\% of queries still finish below reranked dense-only; for RRF the prevalence falls by 46\% and 34\% on SciFact and FiQA and not at all on TREC-COVID. These are net changes in how many queries pay the tax, not repairs of the original failures: on SciFact 27 of the 30 originally hurt queries are no longer hurt after reranking but 17 others become hurt, on FiQA 75 of 95 are repaired and 81 newly hurt, and on TREC-COVID 5 of 7 are repaired and 9 newly hurt (Supplementary Table~S12), so the reranker largely reshuffles which queries pay the tax. On NFCorpus the reranker raises dense-only by only 0.003 while lowering both fused arms (interpolation 0.355$\to$0.350), so the 41\% drop in the prevalence of the deployable tax there (47 of 59 hurt queries repaired, 23 newly hurt) is partly convergence rather than repair. Against the reranked oracle (Supplementary Table~S10), interpolation's tax falls from 20\% to 13\% on SciFact and from 39\% to 28\% on NFCorpus but is essentially unchanged on FiQA (23\% to 24\%). On ArguAna and SciDocs the cross-encoder degrades retrieval (dense-only falls by 0.116 and 0.007), on ArguAna the known weakness of relevance rerankers on counter-argument retrieval~\cite{thakur2021beir}, and on the duplicate-question Quora set it leaves dense-only unchanged ($+0.002$) while lowering the interpolated arm (0.901$\to$0.893); on these three corpora the near-disappearance of the deployable tax (ArguAna 19\%$\to$2\%, Quora 9\%$\to$1\%) is not recovery: the arms converge on the reranker's own ordering, at or below the best unreranked arm, and ArguAna's oracle tax actually rises to 49\%. Two further observations, at depth 30 with this reranker, characterize the second retriever's value in a reranked pipeline. First, once reranked, the candidate source matters little: interpolation candidates score within 0.011 nDCG@10 of dense candidates on six corpora and $+0.018$ on TREC-COVID (paired 95\% bootstrap intervals in Supplementary Table~S12; these are observed differences at depth 30 with one reranker, not a ceiling), and RRF candidates within 0.014 on five, with the exceptions again on the balanced TREC-COVID ($+0.044$) and on ArguAna ($+0.069$, where BM25's candidates suit the reranker better than dense's). Second, where the reranker works, its lift on dense-only exceeds the first-stage gain that fusion delivers over dense-only on three of the five corpora where it does not degrade dense-only (reranking vs.\ fusion: SciFact 0.019 vs.\ 0.011, FiQA 0.042 vs.\ 0.024, TREC-COVID 0.142 vs.\ 0.051; NFCorpus and Quora are the exceptions, 0.003 vs.\ 0.018 and 0.002 vs.\ 0.010). A reranker therefore limits the downstream impact of the tax only on corpora where it is reliable, and that reliability has to be validated per corpus exactly as the fusion decision must be, and on such corpora the second retriever added no more than 0.011 nDCG@10 in our measurements unless sparse and dense are balanced. These results are specific to one cross-encoder (BGE-reranker-v2-m3) at one depth (30) over BGE-small candidates: they are informative about that configuration and should not be read as a general upper bound on what reranking can recover, since a different reranker, a deeper candidate list, or a listwise LLM reranker could recover more of the tax or less of it, and rerankers themselves degrade with pool size~\cite{jacob2025drowning}.

\begin{table*}[!t]
\caption{Reranking recovery (BGE-reranker-v2-m3 over the top-30 of each candidate source, BGE-small, Anserini BM25; Quora subsampled to 1{,}000 queries). ``First stage'' and ``Reranked'' give nDCG@10 before and after reranking; Tax vs.\ dense is the fraction of queries on which the fusion arm scores below the dense arm, before and after reranking, and Net\,$\Delta$ the net change in the prevalence of that tax ($1-\text{after}/\text{before}$); the matched-query transitions behind it are in Supplementary Table~S12. On ArguAna and SciDocs the reranker lowers nDCG@10 for every arm, and on Quora it leaves the dense arm unchanged and lowers the interpolated arm, so the collapse of the tax on these corpora reflects the reranker overriding first-stage order, not recovery.}
\label{tab:rerank}
\centering
{\footnotesize\setlength{\tabcolsep}{3pt}%
\begin{tabular}{l cc ccc ccc ccc}
\toprule
 & \multicolumn{2}{c}{First stage} & \multicolumn{3}{c}{Reranked (nDCG@10)} & \multicolumn{3}{c}{Tax vs.\ dense, Interp(0.7)} & \multicolumn{3}{c}{Tax vs.\ dense, RRF} \\
\cmidrule(lr){2-3}\cmidrule(lr){4-6}\cmidrule(lr){7-9}\cmidrule(lr){10-12}
Dataset & Dense & Interp & Dense & Interp & RRF & Before & After & Net\,$\Delta$\,\% & Before & After & Net\,$\Delta$\,\% \\
\midrule
SciFact & 0.720 & 0.731 & 0.739 & 0.738 & 0.740 & 10.0 & 6.7 & 33 & 15.3 & 8.3 & 46 \\
FiQA & 0.385 & 0.409 & 0.426 & 0.429 & 0.426 & 14.7 & 15.6 & -6 & 30.4 & 19.9 & 34 \\
NFCorpus & 0.337 & 0.355 & 0.340 & 0.350 & 0.354 & 18.3 & 10.8 & 41 & 23.8 & 21.7 & 9 \\
ArguAna & 0.595 & 0.603 & 0.479 & 0.490 & 0.548 & 19.1 & 1.8 & 90 & 32.9 & 4.4 & 87 \\
SciDocs & 0.197 & 0.207 & 0.190 & 0.190 & 0.184 & 20.9 & 21.0 & 0 & 27.6 & 26.8 & 3 \\
TREC-COVID & 0.666 & 0.717 & 0.808 & 0.826 & 0.852 & 14.0 & 22.0 & -57 & 26.0 & 26.0 & 0 \\
Quora & 0.891 & 0.901 & 0.893 & 0.893 & 0.892 & 9.2 & 0.6 & 94 & 17.7 & 1.5 & 92 \\
\bottomrule
\end{tabular}}
\end{table*}

\subsection{Efficiency}
\label{sec:efficiency}
Table~\ref{tab:efficiency} gives the measured cost of each component on the workstation of Section~\ref{sec:setup-rerank} (CPU cores for retrieval and fusion, integrated GPU for the reranker). Adding BM25 to a dense stack is cheap: a Lucene query takes 1.1--4.3\,ms at the median (up to 7\,ms at the 95th percentile), the inverted index occupies 8--266\,MB, and indexing the 171{,}000 TREC-COVID abstracts takes 11.5\,s with eight threads (title and body as separate fields, the configuration of the prebuilt indexes used for retrieval). Adding a dense retriever to a lexical stack is where the cost lies: encoding the query takes 5--7\,ms for BGE-small (19\,ms on ArguAna, whose queries are full arguments), exact search over the embedding matrix 0.1--12\,ms, and the matrix itself 6--803\,MB for BGE-small and 15--2{,}142\,MB for BGE-large; encoding the corpus runs at 40 abstracts per second for BGE-small and 5 for BGE-large on CPU (784 and 83 per second on Quora's short questions; throughput measured on 512 documents for the dense encoders and 128 for SPLADE), so by extrapolation indexing TREC-COVID costs 71\,min and 9.7\,h, respectively, against Lucene's 11.5\,s. SPLADE sits in between: 10 abstracts per second to index and 17--19\,ms to encode a query (53\,ms on ArguAna). The fusion step itself is negligible (under 0.1\,ms). For comparison, the cross-encoder of Section~\ref{sec:rerank-results} scored 19 pairs per second in 16-bit precision on the same machine's integrated GPU, i.e.\ about 1.6\,s per query at depth 30, two to three orders of magnitude more than the second retriever. The second retriever is therefore the cheap lever and the reranker the expensive one; what the second retriever buys in quality is the subject of Sections~\ref{sec:fusion-tax-results} and~\ref{sec:rerank-results}.

\begin{table*}[!t]
\caption{Cost of each component on one consumer workstation (Apple M5, CPU cores; no discrete GPU): index sizes and per-query latencies, median over 200 queries (50 on TREC-COVID, and 50 on every corpus for SPLADE) at batch size one. Dense enc.\ and FAISS are BGE-small query encoding and exact search; SPLADE enc.\ is query encoding only; Interp and RRF are the fusion step over two top-100 lists.}
\label{tab:efficiency}
\centering
{\footnotesize\setlength{\tabcolsep}{3pt}%
\begin{tabular}{l r rrr rrr r rr}
\toprule
 & & \multicolumn{3}{c}{Index size (MB)} & \multicolumn{6}{c}{Per-query latency, median (ms)} \\
\cmidrule(lr){3-5}\cmidrule(lr){6-11}
Dataset & Docs & Lucene & BGE-sm & BGE-lg & BM25 & Dense enc. & FAISS & SPLADE enc. & Interp & RRF \\
\midrule
SciFact & 5k & 11 & 8 & 21 & 1.7 & 6.6 & 0.1 & 17.8 & 0.08 & 0.05 \\
FiQA & 58k & 67 & 88 & 236 & 2.0 & 6.1 & 1.3 & 18.1 & 0.08 & 0.06 \\
NFCorpus & 4k & 8 & 6 & 15 & 1.1 & 5.3 & 0.1 & 18.5 & 0.07 & 0.05 \\
ArguAna & 9k & 13 & 13 & 36 & 2.9 & 19.1 & 0.2 & 52.9 & 0.08 & 0.06 \\
SciDocs & 26k & 238 & 39 & 105 & 2.7 & 6.2 & 0.6 & 17.1 & 0.08 & 0.05 \\
TREC-COVID & 171k & 266 & 263 & 702 & 4.3 & 6.3 & 3.9 & 18.4 & 0.08 & 0.06 \\
Quora & 523k & 69 & 803 & 2142 & 1.7 & 5.7 & 11.8 & 18.4 & 0.08 & 0.06 \\
\bottomrule
\end{tabular}}
\end{table*}

\subsection{Variance Decomposition}
A one-way ANOVA states an obvious but useful point: dataset identity explains far more nDCG@10 variance than the fusion method ($\eta^2_{\text{dataset}}=0.40$ vs.\ $\eta^2_{\text{method}}=0.01$), largely because baseline difficulty ranges from 0.20 (SciDocs) to 0.89 (Quora). This is not a claim that the fusion choice is unimportant; it is the reason we report every comparison per corpus rather than pooled, and it is consistent with the corpus-specificity of fusion failure above.

\section{Discussion}
\label{sec:discussion}

\subsection{RRF Is Not a Safe Default, and When It Is the Right Choice}
Our results qualify the default recommendation of RRF that the documentation surveyed in Section~\ref{sec:related-fusion} carried at the time of writing. On the five corpora where the dense retriever is clearly the stronger component, dense-weighted interpolation beats RRF in aggregate (significantly on four), RRF's per-query tax against the oracle is higher (20--40\% vs.\ 13--36\%), and RRF's net change against simply using the dense retriever is negative. RRF discards score magnitude, the information that protects the documents the stronger component ranked highly (a document ranked first by a confident retriever counts the same as one ranked first by a hair), but it also weights the two components equally where the interpolation it loses to weights dense 0.7. The two effects can be separated: a weighted RRF with the same 0.7/0.3 weights and no score magnitudes closes 25--73\% of the RRF-to-interpolation gap on the five dense-led corpora (Supplementary Table~S13), so weighting accounts for roughly half of RRF's disadvantage and the score information for the rest. The picture reverses where sparse and dense are balanced: on TREC-COVID, RRF and CombMNZ beat dense-heavy interpolation by 0.08--0.09 nDCG@10 and help 70--74\% of queries, and on NFCorpus the three methods tie. This is not a contradiction but the regime structure that the calibration literature predicts (Section~\ref{sec:related-calibration}): interpolation is the better tool when there is a confidence worth keeping, rank fusion when there is not. The practical consequence is that the fusion method should follow the measured sparse--dense balance on the target corpus rather than a global default, and that the balance can be measured on a labeled sample of a few hundred queries before choosing (Section~\ref{sec:framework}).

\subsection{When Does Fusion Help?}
The quadrant decomposition (Table~\ref{tab:quadrant-tax}) makes the mechanism explicit. Fusion helps when the two retrievers are complementary \emph{and} comparably reliable on a query: in Both-Win the two rankings can promote different relevant documents, although Table~\ref{tab:quadrant-tax} shows that the reordering runs both ways there and nearly cancels, and in Both-Fail fusion can only pull a relevant document up from below the cutoff. Fusion hurts when one retriever is decisively better and the other contributes noise: in Dense-Only and Sparse-Only, fusion dilutes the strong signal and demotes documents the better retriever had already surfaced, and a dense-heavy weight sacrifices the Sparse-Only minority almost entirely (87--95\% hurt). Because a stronger dense encoder widens its lead on the queries it already wins without shrinking the off-diagonal share, the tax does not decline along the ladder. The value of fusion on a corpus is thus governed by the balance and complementarity of its two retrievers on that corpus's query distribution, not by the absolute quality of either.

\subsection{Corpus Patterns in Brief}
SciFact (precise scientific terminology) has the cleanest profile: 76\% Both-Win and the lowest tax after Quora. FiQA is among the hardest, with 30\% Both-Fail and a 23\% Dense-Only share where only semantics help. NFCorpus carries the highest tax outside the 50-query TREC-COVID set (39\%) because heavy synonymy gives both retrievers partially relevant documents that fusion frequently reorders. ArguAna is dense-dominated (20\% Dense-Only) and pays a 30\% tax even though, with the standard BM25, fusion matches dense-only in aggregate. SciDocs resembles a harder SciFact, with a third of its queries paying the tax. TREC-COVID (50 queries, 100\% Both-Win, balanced retrievers) is the regime in which rank fusion wins and fusion helps most queries, yet it still pays the highest oracle tax because when both retrievers rank a relevant document first, fusion can only displace it. Quora (duplicate questions) is dense-dominated and accurate, so fusion has the least to add and incurs the lowest tax (13\%), yet even there it demotes relevant documents on one query in eight.

\subsection{Practical Decision Framework}
\label{sec:framework}
\begin{enumerate}
    \item \textbf{Measure the balance first.} On a labeled sample of the target corpus, compare BM25 and dense nDCG@10 query by query. Draw queries randomly from the target query stream, using proportional stratification across known traffic segments; if rare segments are oversampled, weight the overall estimate back to traffic proportions and report segment results separately. Judge at least the pooled top-10 of both retrievers, extending the pool to any fusion candidates being evaluated. As a planning illustration, the paired dense-minus-BM25 nDCG@10 differences in our data have standard deviations of 0.17--0.36. For independent queries with similar variability, the approximate 95\% confidence-interval half-width is $1.96s/\sqrt{n}$, where $s$ is the sample standard deviation: about 0.03--0.07 at 100 queries, 0.02--0.05 at 200, and 0.016--0.035 at 400. A practical starting point is therefore 200--400 representative queries, with a larger sample for smaller gaps, heterogeneous traffic, or narrower precision requirements. These are planning values, not a guaranteed sample size for a new corpus; estimate its variability in a pilot and pre-specify the required precision. An interval overlapping zero does not establish equivalence. Select the weight by cross-validation (Section~\ref{sec:alpha}) and measure the tax on held-out queries rather than on the queries used to select it. If dense leads clearly (in our data by 0.04 nDCG@10 or more, against 0.01 on the balanced corpora), use dense-weighted interpolation with $\alpha\approx0.6$--$0.7$; this weight transferred across corpora at a cost of at most 0.014 nDCG@10 except on TREC-COVID, whose optimum lies at 0.5 (Table~\ref{tab:alpha}). If the two are balanced, use RRF or CombMNZ, or interpolation near $\alpha=0.5$.
    \item \textbf{Treat dense-only as a live option.} Fusion beat dense-only in aggregate on every corpus with the primary encoder, and on 33 of the 35 encoder--corpus pairs overall (the exceptions are both Qwen3-Embedding: SciDocs by 0.003 and ArguAna by 0.036), but by only 0.008--0.051 nDCG@10, while 8--21\% of queries lost 0.04--0.21 nDCG@10 on average. For a rank-sensitive consumer with a loss-averse objective, the risk-sensitive tests favor dense-only on six of the seven corpora.
    \item \textbf{Do not expect a stronger or newer encoder to remove the tax.} It persisted across a 22M--596M ladder and three model generations; the decision to fuse should be revisited per corpus, not settled once.
    \item \textbf{Prefer a per-corpus static policy to per-query routing on cheap features.} The failure signal does not transfer across corpora and its features change meaning with the sparse--dense balance (Section~\ref{sec:feature-results}).
    \item \textbf{Budget the second retriever against the reranker.} Adding BM25 to a dense stack costs 1.1--4.3\,ms per query and a small index, whereas a cross-encoder over 30 candidates costs about 1.6\,s per query on the same hardware (Table~\ref{tab:efficiency}). Fusion with a balance-keyed weight is the cheap lever where no reranker is affordable; where a reliable reranker is already in place, the second retriever added at most 0.011 nDCG@10 in our measurements except on balanced corpora, and the budget is better spent on validating the reranker per corpus.
\end{enumerate}

\subsection{Why Per-Query Routing Is Hard}
A natural response to the tax is to fuse only when fusion is predicted to help. Section~\ref{sec:feature-results} shows why this is harder than it looks: the failure signal exists within a corpus but its carriers and their directions change across corpora, so a router calibrated on one corpus can be at or below chance on another, the practitioner has no cheap way to know which, and the held-out router of Supplementary Table~S11 confirms that acting on the signal does not pay. This is consistent with QPP-weighted fusion working best when tuned to a target distribution~\cite{santra2026qpp} and with the observation that classical QPP degrades on neural rankers~\cite{faggioli2023qppneural}. It also frames the per-query gap that routing would have to close: the oracle gains 0.02--0.05 nDCG@10 over static fusion on every corpus, which is worth pursuing, but the evidence here says the signal must come from richer, per-corpus, or model-based sources rather than transferable heuristics.

\subsection{Temporal Stability as Encoders Evolve}
The ladder spans encoder generations from 2021 (MiniLM) through 2023 (BGE v1.5) to 2025 (Qwen3-Embedding), and the 2025 model pays a tax of 14--43\%, a range overlapping those of the 2021 and 2023 models, despite being the strongest encoder in the study on four corpora. Two structural reasons suggest the tax will outlive the next encoder generation as well: the off-diagonal quadrants are defined by disagreement between two retrievers, which a better dense model reduces only on the dense side, and score interpolation must still normalize two incomparable distributions, which no encoder improvement changes. What a newer encoder does change is the sparse--dense balance, and hence which fusion method and weight are appropriate, so the balance should be re-measured whenever the encoder is updated.

\subsection{Limitations and Future Work}
(1)~We evaluate first-stage retrieval and one reranking stage, not answer generation; the reranking experiment measures the exposure of a retrieve-then-rerank pipeline at one depth with one reranker, and the effect on generated answers remains open. (2)~The reranking results are for one cross-encoder applied to the top-30 candidates of each source; they are not a general upper bound on reranking recovery, since deeper reranking or a different reranker could recover more of the tax, or less, as rerankers themselves degrade with pool size~\cite{jacob2025drowning}. (3)~TREC-COVID has 50 queries, so its estimates carry wide intervals, and the multi-million-document BEIR corpora are omitted under a CPU budget. (4)~Efficiency was measured on one consumer workstation without a discrete GPU; absolute latencies will differ on production hardware, and so will the ratios between components, since a GPU shortens dense encoding and reranking but not Lucene; what we expect to survive is the ordering, with BM25 the cheapest component per query and the cross-encoder the most expensive. (5)~Our predictability analysis covers cheap QPP features and a dense query embedding. An LLM-based effectiveness scorer of the kind DAT~\cite{hsu2025dat} uses to set its per-query weight may carry a signal that transfers better; the natural next step, left to future work, is to score each component's top-ranked documents with an LLM judge as DAT does, add the two scores to the feature set of Section~\ref{sec:features}, and test transfer under the same leave-one-corpus-out protocol while reporting the judge's per-query cost. (6)~Two BEIR artifacts apply identically to all methods and slightly depress absolute scores: five ArguAna queries whose sole relevant document is absent from the corpus, and exact-duplicate documents under different identifiers.

\section{Conclusion}
\label{sec:conclusion}
We have given a per-query accounting of hybrid retrieval that treats ``do not fuse'' as a first-class outcome. With a standard Anserini BM25, dense encoders from 22M to 596M parameters, and a learned-sparse alternative, the best fusion method beats the deployable dense-only retriever in aggregate on every corpus with the primary encoder and on 33 of the 35 encoder--corpus pairs, yet demotes relevant documents below the better component on 13--48\% of queries and below dense-only on 8--21\%, with losses on a hurt query of the same order as the gains on a helped one and significant downside risk under loss-averse evaluation on six of the seven corpora. The tax is mostly a ranking effect (recall@10 is preserved in aggregate, although a minority of the hurt queries lose a relevant document from the top-10), it is most likely on the queries that only one retriever covers (though on corpora where both retrievers usually succeed, most hurt queries are ones both cover), and it persists across the encoder ladder and the SPLADE substitution; the weight that maximizes aggregate nDCG@10 transfers across corpora under held-out selection, while cheap per-query routing does not, because the features that predict failure change meaning with the sparse--dense balance. A cross-encoder second stage removes part of the tax only where it is itself reliable, overwrites it where it is not, and leaves interpolated and dense candidate sets within 0.011 nDCG@10 of each other once reranked with the one reranker and depth tested, except where sparse and dense are balanced. The practical guidance follows the mechanism: measure the sparse--dense balance on the target corpus, choose dense-weighted interpolation where dense leads and rank fusion where the two are even, keep dense-only on the table for rank-sensitive consumers, and re-check the decision when the corpus or the encoder changes.

\section*{Acknowledgment}
In accordance with IEEE's policy on the disclosure of artificial intelligence (AI) use, the authors state that the AI assistant Claude (Anthropic) was used to check and refine the language of the author-written manuscript text. The research questions, methodology, experimental data, analyses, and conclusions are the authors' own, and all content was reviewed and verified by the authors, who take full responsibility for the accuracy and integrity of the article.

\section*{Data Availability}
All seven evaluation datasets are publicly available BEIR benchmarks~\cite{thakur2021beir}, and the BM25 runs use the official prebuilt Anserini indexes~\cite{kamalloo2024brewingbeir}. The complete pipeline (retrieval, fusion, evaluation, diagnostics, held-out weight selection, reranking, efficiency measurement, and every analysis and table script), the summary result files from which all tables and figures derive, and a step-by-step reproducibility guide (\texttt{REPRODUCE.md}) that traces each reported number to its source data are provided as supplementary material accompanying this article and, together with the per-query retrieval runs of every retriever on every corpus, are archived on Zenodo (\url{https://doi.org/10.5281/zenodo.22298900}; the version reviewed here is \url{https://doi.org/10.5281/zenodo.22298901}).

\bibliographystyle{ieeetr}
\phantomsection
\bibliography{references}

\begin{IEEEbiography}[{\includegraphics[width=1in,height=1.25in]{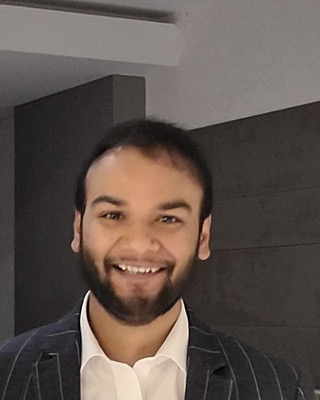}}]{Aaditya Chauhan}
is a researcher and engineer specializing in information retrieval, retrieval-augmented generation, and large-scale search systems. With over a decade of experience in distributed systems and production search, he has designed and operated consumer-scale retrieval infrastructure spanning hybrid retrieval, vector search, dense retrieval compression, and embedding lifecycle management. His research centers on evaluation methodology and reproducible benchmarking for retrieval systems, with a focus on the real-world trade-offs between storage, latency, and relevance in dense and hybrid retrieval. He serves as a reviewer for several ACM conferences and for IEEE Access, and has served as a session chair at the International Conference on Enterprise Information Systems (ICEIS).
\end{IEEEbiography}

\begin{IEEEbiography}[{\includegraphics[width=1in,height=1.25in]{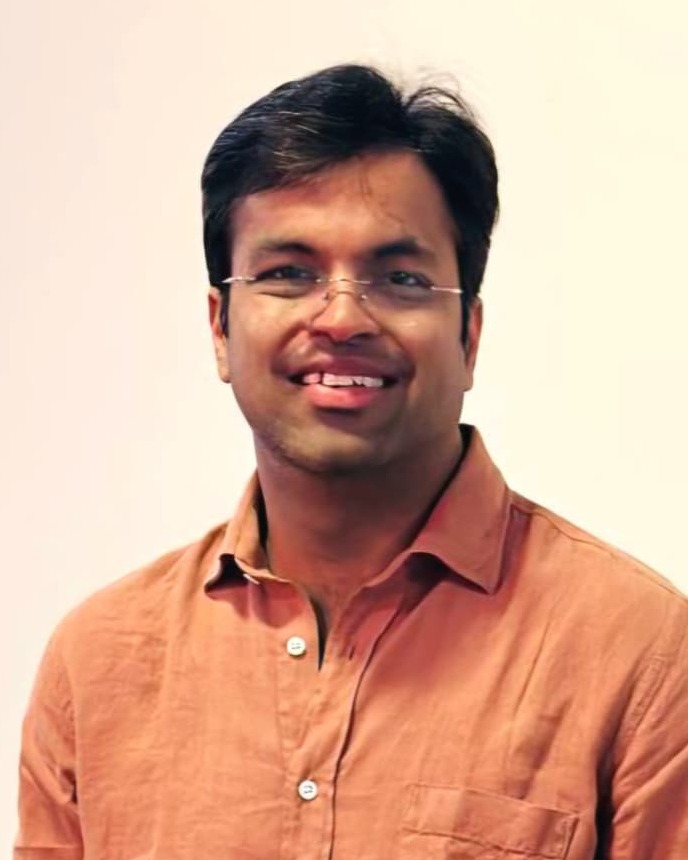}}]{Arun Agarwal}
is an engineering leader and practitioner-researcher working at the intersection of information retrieval, on-device machine learning, and large-scale systems engineering. His research interests include retrieval-augmented generation, embedding compression and index efficiency, and evaluation methodology for retrieval systems under real-world storage, latency, and cost constraints. He is particularly interested in building reproducible experimental frameworks that connect research methods to production deployment decisions, drawing on his experience leading multi-team software organizations that ship media experiences across consumer device platforms.
\end{IEEEbiography}

\EOD

\end{document}